\documentclass[preprint,journal]{vgtc}       % preprint (journal style)

\ifpdf%                                % if we use pdflatex
  \pdfoutput=1\relax                   % create PDFs from pdfLaTeX
  \usepackage{graphicx}                % allow us to embed graphics files
  \DeclareGraphicsExtensions{.pdf,.png,.jpg,.jpeg} % for pdflatex we expect .pdf, .png, or .jpg files
\else%                                 % else we use pure latex
  \usepackage{graphicx}                % allow us to embed graphics files
  \DeclareGraphicsExtensions{.eps}     % for pure latex we expect eps files
\fi%

\graphicspath{{figures/}{pictures/}{images/}{./}} % where to search for the images

\usepackage{microtype}                 % use micro-typography (slightly more compact, better to read)
\PassOptionsToPackage{warn}{textcomp}  % to address font issues with \textrightarrow
\usepackage{textcomp}                  % use better special symbols
\usepackage{mathptmx}                  % use matching math font
\usepackage{times}                     % we use Times as the main font
\usepackage{cite}                      % needed to automatically sort the references
\usepackage{tabu}                      % only used for the table example
\usepackage{booktabs}                  % only used for the table example

\usepackage{xcolor}
\usepackage[linesnumbered,ruled,vlined]{algorithm2e}
\usepackage{amsmath}

\makeatletter
\usepackage{tikz}
\newcommand*\circled[2][1.6]{\tikz[baseline=(char.base)]{
    \node[shape=circle, draw=black!70, fill=gray!20, inner sep=0pt, 
        minimum size={2pt},] (char) {\vphantom{WAH1g}#2};}}
\makeatother

\newcommand*\circledtwo[3][1.6]{\tikz[baseline=(char.base)]{
    \node[shape=circle, draw=black!70, fill=gray!20, inner sep=0pt, 
        minimum size={2pt},] (char) {{\textcolor{black}{\vphantom{WAH1g}#2}}{\tiny\textcolor{black}{\vphantom{WAH1g}#3}\par}};}}
\makeatother

\onlineid{1307}

\vgtccategory{Research}
\vgtcpapertype{system}

\newcommand\systemname{\textit{VATLD}}
\newcommand\tilescape{\textit{TileScape}}
\newcommand\hpcp{\textit{hPCP}}

\newcommand{\inlinep}[1]{\raisebox{-.25\height}{\includegraphics[height=1.5\fontcharht\font`\B]{#1}}}
\mathchardef\mhyphen="2D

\newcommand{\revisionblue}[1]{\textcolor{black}{#1}}

\title{\systemname: A \textit{{V}}isual \textit{{A}}nalytics System to Assess, Understand and Improve \textit{{T}}raffic \textit{{L}}ight \textit{{D}}etection}

\author{Liang Gou, Lincan Zou, Nanxiang Li, Michael Hofmann, Arvind Kumar Shekar, Axel Wendt and Liu Ren}
\authorfooter{
  \vspace{-1em}
\item
 \{Liang Gou, Lincan Zou, Nanxiang Li, Liu Ren\} are with Robert Bosch Research and Technology Center, USA. E-mails: \{liang.gou, lincan.zou, nanxiang.li, liu.ren\}@us.bosch.com.
\item
 \{Michael Hofmann, Arvind Kumar Shekar, Axel Wendt\} are with Robert Bosch GmbH, Germany. E-mails: \{michael.hofmann11, arvindkumar.shekar, axel.wendt\}@de.bosch.com
}

\shortauthortitle{Biv \MakeLowercase{\textit{et al.}}: Global Illumination for Fun and Profit}
\abstract{Traffic light detection is crucial for environment perception and decision-making in autonomous driving. State-of-the-art detectors are built upon deep Convolutional Neural Networks (CNNs) and have exhibited promising performance. However, one looming concern with CNN based detectors is how to thoroughly evaluate the performance of accuracy and robustness before they can be deployed to autonomous vehicles. In this work, we propose a visual analytics system, \systemname, equipped with a disentangled representation learning and semantic adversarial learning, to assess, understand, and improve the accuracy and robustness of traffic light detectors in autonomous driving applications. The disentangled representation learning extracts data semantics to augment human cognition with human-friendly visual summarization, and the semantic adversarial learning efficiently exposes interpretable robustness risks and enables minimal human interaction for actionable insights. We also demonstrate the effectiveness of various performance improvement strategies derived from actionable insights with our visual analytics system, \systemname, and illustrate some practical implications for safety-critical applications in autonomous driving.

} % end of abstract


\CCScatlist{ % not used in journal version
 \CCScat{K.6.1}{Management of Computing and Information Systems}%
{Project and People Management}{Life Cycle};
 \CCScat{K.7.m}{The Computing Profession}{Miscellaneous}{Ethics}
}

\teaser{
  \centering
  \includegraphics[width=0.8\linewidth]{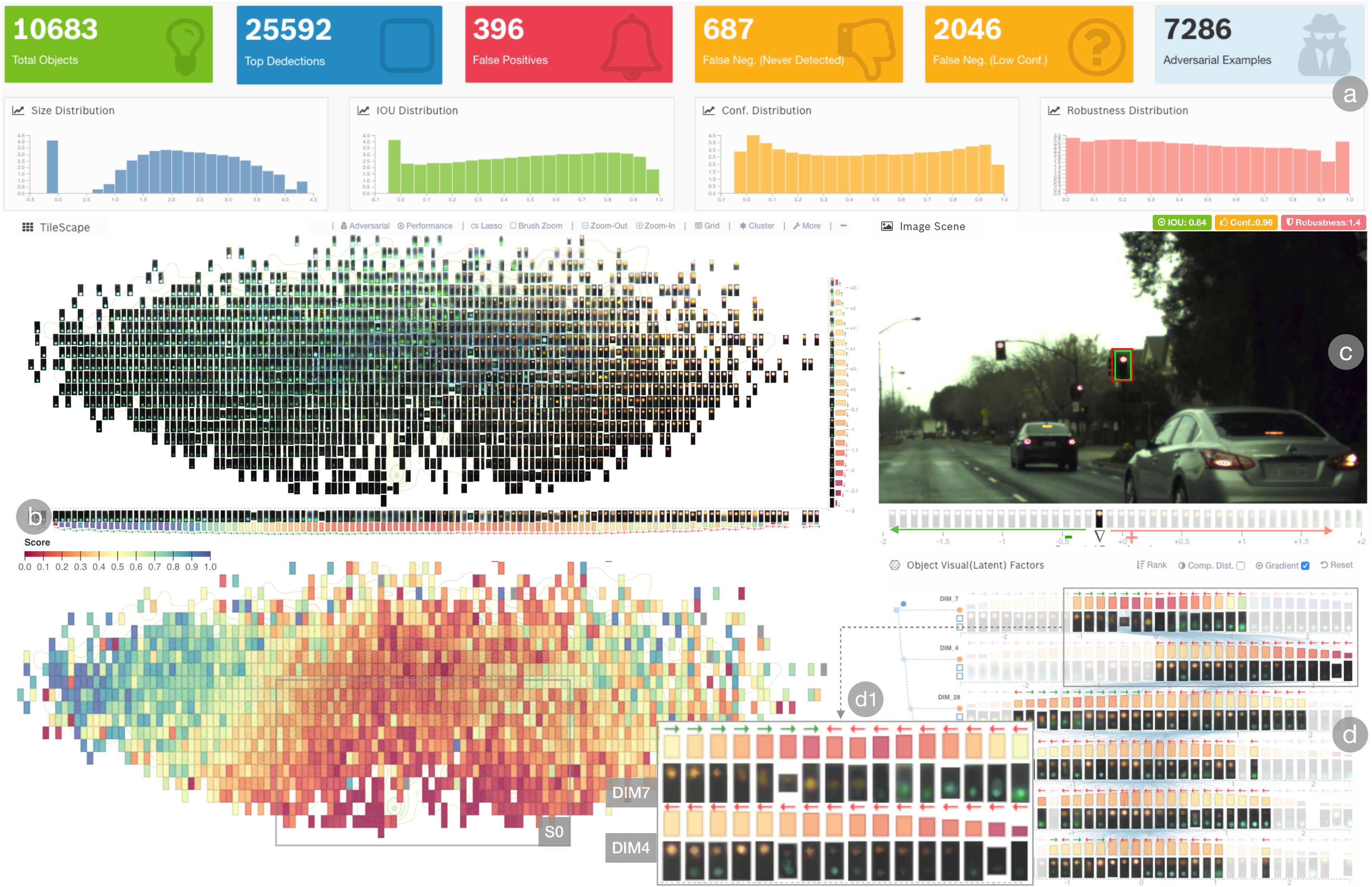}
   \vspace{-1em}
  \caption{\revisionblue{The \systemname\ user interface: (a) Summary and navigation of key performance statistics; (b) Visual landscapes of traffic lights (upper) and performance scores (lower) over the first two PCA components of semantic dimensions; (c) A live view to detect a traffic light from either real data or adversarial; (d) Ranked latent dimensions with information of semantics, performance and gradients.} }
  
  \label{fig:teaser}
}
\vgtcinsertpkg

\begin{document}



\firstsection{Introduction}

\maketitle

%% \section{Introduction} %for journal use above \firstsection{..} instead

Traffic light detection is an essential component in autonomous driving. It helps autonomous cars perceive driving environments by locating relevant traffic lights and also supports cars to make right decisions by recognizing the status of lights. The state-of-the-art traffic light detectors largely rely on deep Convolutional Neural Networks (CNNs) that have exhibited superior performance in many computer vision tasks such as image classification, object detection, semantic segmentation and so on. These detectors are usually trained upon general purpose object detectors (such as SSD \cite{Liu2015}, YOLO \cite{Redmon2016} and Faster R-CNN \cite{Ren2017} locating and recognizing animals, person, vehicles, etc \cite{Lin2015}) and then fine tuned with domain-specific data (driving scenes with traffic lights)\cite{Behrendt2017} or combined with other prior knowledge about driving scenes, such as object distribution in a scene \cite{Possatti}.  

Despite the promising results of CNN based detectors, one concern is how to thoroughly assess, understand and further improve detector performance before they can be deployed to autonomous vehicles. The concern is two-fold: a) \textbf{\textit{accuracy}} evaluation and improvement over massive \textit{acquired data} (training and testing data); b) \textbf{\textit{robustness}} evaluation and improvement over \textit{unseen data} (potential vulnerability). 

Firstly, it is a non-trivial task to assess model accuracy and understand when and why detectors tend to fail. The current evaluation and benchmark methods of model accuracy heavily rely on aggregated and over-simplified metrics, such as \textit{mAP} (mean Average Precision) \cite{Lin2015}, and fail to provide interpretable and contextual information to understand model performance. 
% In practice, model developers are eager to see what object features and driving environments may cause different failures of detectors, such as different illumination settings (sun glare or dark in the tree), or object sizes in the scene (at different distances), or interference with confusing objects (similar but irrelevant objects). 
Furthermore, although rising attention has been paid to the explainability of general CNNs \cite{LiuCNNVis, Liu2017, Choo2018, Bau2017, Ribeiro2016LIME, Wang2018, Wang2019, Bilal2018} , the method of unveiling how CNN based detectors perform still lack of investigation. During a model building phase, interactive tools are desirable to support model assessing, understanding and improvement instead of a single accuracy number. 

Another burning need is to identify a detector's potential vulnerability, and then assess and improve the robustness over potential vulnerable cases. Recently, the advance of adversarial attack and robustness research bears much potential to reveal the vulnerability in DNN (Deep Neural Networks) \cite{Szegedy2014, Goodfellow2015, Eyas2018, Li-NATTACK}. Adversarial machine learning fools a classifier with small perturbation of inputs with the gradient information obtained from DNNs. 

\revisionblue{However, two challenges are on the horizon by applying current adversarial attack methods to understand, evaluate, and improve the robustness of detectors. First, most adversarial attack methods do not generate examples with meaningful changes. These methods aim to fool target models by adding imperceivable noises \cite{Szegedy2014, Goodfellow2015, Eyas2018, Li-NATTACK}, and therefore these noises do not provide physical meanings or semantics for human to understand model robustness, and also provide little guidance to improve robustness in the physical world. Also, the mechanism understanding adversarial landscape and improving robustness of a model is desirable. }%nothing short of scarce. 
For example, with current methods, we do not know what the common patterns of learned adversarial examples are, why they exist, and how to improve.

All told, to asses, understand and improve the performance of traffic light detectors, we need to overcome two hurdles of dissecting model accuracy over existing data, and also assessing and improving model robustness over unseen cases. 

Aiming at above challenges, we propose a visual analytics system, \systemname, to assess, understand, and improve the accuracy and robustness of traffic light detectors in autonomous driving applications. Specifically, in this work, our contributions include:

\vspace{-0.6em}
\begin{itemize}
\itemsep-0.3em 
\item offering a novel visual analytics method to asses, understand and improve a detector's performance of both accuracy and robustness, guided by a semantic representation learning and a minimal human-in-the-loop approach.

\item adapting a representation learning method to efficiently summarize, navigate and diagnose the performance of detectors over large amounts of data. This method extracts low-dimensional representation (i.e. latent space) with disentangled intrinsic attributes (such as color, darkness, background etc.) of traffic lights and serves a fundamental representation for both human-friendly visualization and semantic adversarial learning. 

\item proposing a new semantic adversarial learning algorithm to efficiently reveal the robustness issues of detectors without knowing detector parameters. The adversarial learning efficiently generates unseen test cases by searching the learned latent space, and the adversarial results are also interpretable with meaningful perturbation in the latent space. 

%\item coining a novel robustness metric to assess and interpret these risks based on the adversarial learning method, and also providing interactive visualization to analyze the robustness landscape of a detector. 

\item demonstrating the effectiveness of model improvement strategies guided by visual analytics for real world problems, and illustrating a promising way of injecting human intelligence into models with few human interactions.

\end{itemize}
\vspace{-0.6em}
\revisionblue{In sum, we present, \systemname, as the first research effort using a visual analytics approach to alleviate some trustworthy AI issues in autonomous driving applications. We hope this work can raise some discussion and open research opportunities for applying visual analytics and human-in-the-loop approaches to the domain of autonomous driving.}

\section{Related Work}
\subsection{Object Detection and Traffic Light Detection}
Object detection is a fundamental computer vision task to locate and recognize the instances of visual objects from a certain class (such as humans, animals, vehicles, or others) in images. Riding the recent wave of deep learning, generic object detectors with state-of-the-art performance are mostly built upon powerful convolutional neural networks (CNN). CNN based detectors could be roughly categorized into two groups: two-stage and one-stage detectors. Two-stage detectors, such as R-CNN (``Region-based CNN") based detectors\cite{Girshick2014, Girshick2015, Ren2017}, first propose a set of regions of interests (bounding boxes) by a searching method or a regional proposal network, and then use a classifier to recognize these regions. By contrast, one-stage detectors directly search and classify possible locations over one feature map (e.g. YOLO \cite{Redmon2016}) or multi-scale feature maps (e.g. SSD \cite{Liu2015}, RetinaNet \cite{Lin2017}). Although these generic object detectors have achieved promising performance over multiple data-sets (e.g. coco dataset \cite{Lin2015}), specialized detectors are desired for various perception tasks in autonomous driving applications, such as detection for traffic lights, lanes, and pedestrians.

Traffic light detector is a specialized detector locating traffic lights and recognizing their status (such as red, green and yellow). These detectors are usually built upon CNN based detectors and fine tuned with domain-specific data (driving scenes with traffic lights)\cite{Behrendt2017}, or combined prior knowledge about driving scenes, such as object distribution \cite{Possatti}, and map information \cite{John2014}.

\subsection{Assessing and Interpreting CNN based Detectors}
A plethora of work has been proposed to assess and interpret generic CNN models in both machine learning and visual analytics communities. These works include model-specific and model-agnostic approaches. Model-specific methods attempt to reveal multiple levels of details learned in CNNs, such as neuron activation and filters \cite{LiuCNNVis}, feature maps \cite{Szegedy2014}, concepts and feature attributions \cite{Bau2017}, while model-agnostic approaches employ intrinsic explainable models, such as decision trees, decision rules \cite{Ming2019}, linear models \cite{Ribeiro2016LIME, Wang2019}, to mimic the prediction behaviors of CNNs. Surveys from both machine learning \cite{Zhang2018} and visual analytics \cite{Liu2017, Choo2018, yuan2021survey} offer more insights into this. 

However, assessing and improving the performance of CNN based detectors, especially for traffic lights, is of scarce. Hoiem et al. \cite{Hoiem} conduct error analysis for generic object detectors by examining the influences of object characteristics (e.g. occlusion, size, aspect ratio, viewpoint etc.) on detection performance. However, these object characteristics are handcrafted and hard to scale. Vondrick et al. \cite{Vondrick2016} present a method to visualize detector features as natural images, but focusing on traditional histogram of oriented gradients (HOG) based not CNN-based detectors. In short, these approaches lack of an intrinsic representation of objects to augment human's cognition for understanding and diagnosing model performance. Moreover, current methods are incapable of revealing and assessing the potential weakness of detectors to ensure the level of confidence that autonomous driving requires.  

\subsection{Adversarial Robustness}
To tackle down the issue of potential weakness for CNN based detectors, recent developments in adversarial robustness research pointed some promising directions. The seminal work on adversarial attack \cite{Szegedy2014} offers us an efficient way to fail a classifier with minimal perturbation of inputs by following the ascending gradient direction of a model. According to the way of obtaining gradient information, adversarial learning methods can be grouped into white-box attacks \cite{Szegedy2014,Goodfellow2015}, using model parameters to calculate gradients, and black-box attacks \cite{Eyas2018,Li-NATTACK}, with limited model queries to estimate gradients. In both approaches, model robustness can be measured by the minimal perturbation of the image pixels needed to fail the model. 

However, most adversarial attack methods do not generate examples with semantic meanings and limit their capability to generate actionable guidance for the model improvement over potential risks in the real world. \revisionblue{Some initial research \cite{Kurakin2017, Chen2019} is conducted on the attacks in the physical world but it is hard to scale because of the scarce of available data with physical perturbation.} Also, we lack of approaches for understanding and interpreting the semantics of adversarial examples. For example, it is not clear what the common patterns of the learned adversarial examples are and if these patterns are explainable or helpful to generate actionable insights in practice.    

In summary, to assess, understand and improve the performance of traffic light detectors, we aim to bridge the research gaps of dissecting the model accuracy over existing data, and also assessing and improving model robustness over unseen vulnerable cases in this work.
%\cite{Cao, Ma2020}

\section{Background and Motivation}
\subsection{How a deep CNN based detector works}
\begin{figure}[tb]
 \centering % avoid the use of \begin{center}...\end{center} and use \centering instead (more compact)
 \includegraphics[width=\columnwidth]{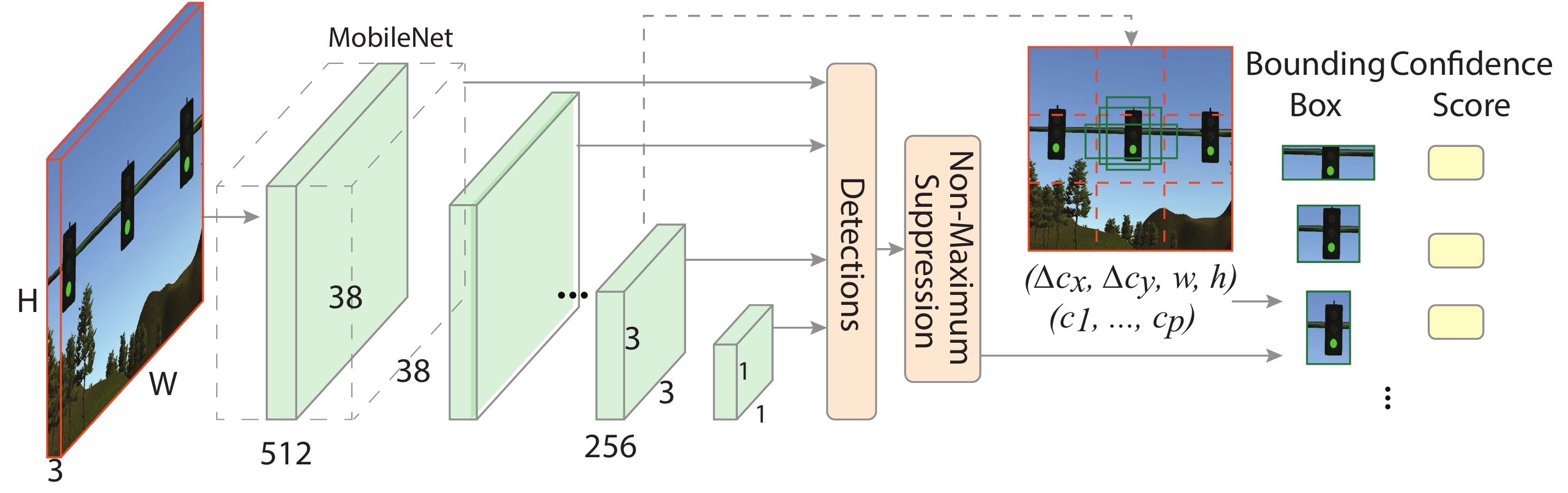}
 \vspace{-2em}
 \caption{A Single Shot multibox Detector (SSD) \cite{Liu2015} for traffic lights.}
 \vspace{-1em}
 \label{fig:ssd}
\end{figure}

In this work, we use a Single Shot multiBox Detector (SSD) \cite{Liu2015} as an example because of its fast detection and high accuracy. 
%Let's see how a SSD works for traffic light detection. 
% ($X_{nn}$ with height $n$ and width $n$)
With an image of driving scene as input, a SSD has two tasks: \textbf{locating} possible objects (traffic lights in this case) with a set of bounding boxes ($\{b_i\} \colon (\Delta c_x, \Delta c_y, w, h$), $\Delta c_x, \Delta c_y$ as the offsets to center coordinates, and $w, h$ as box width and height), and \textbf{recognizing} object categories (red, green, yellow, off, and non-object/background) with confidence scores ($\{c_i\}, i$ as a category) for each predicted box, as shown in \autoref{fig:ssd}. 

A SSD first passes an image thorough a back-bone CNN, such as ResNet \cite{He2016} and MobileNet\cite{Howard2017}, to extract base features (e.g. $38\times38$ with 512 channels in \autoref{fig:ssd}). These features are then converted into smaller size feature maps at different scales. For each cell on a feature map (e.g., a $3\times3$ with depth 256 has $3\times3 = 9$ cells), it makes $k$ predictions of boxes with various aspect ratios, and $p$ class scores for each box with a fixed $3\times3$ convolutional predictors. 
%In the illustrated example, a $5\times5\times512$ feature map yields $5\times5\times k\times (p+4)$ predictions (4 for the bounding box $\Delta c_x, \Delta c_y, w, h$).
This prediction is applied to all feature maps at multiple scales to detect objects with different sizes. Finally, predictions with overlapping boxes are filtered by a non-maximum suppression method, and the outputs are ranked by maximal confidence scores of non-background classes.

To simplify the discussion in the paper, let's consider detector recognition to be a binary classification of traffic-light and non-traffic light (background), $c(x)=c_{traffic light}$. However, this is not a general limitation as the proposed method is model-agnostic and can thus be applied to any object detection network with multiple classes.

%https://medium.com/@jonathan_hui/ssd-object-detection-single-shot-multibox-detector-for-real-time-processing-9bd8deac0e06

\subsection{Metrics for Detector Accuracy }
\textbf{Average Precision (AP)} is a well-adopted metric to evaluate detector accuracy by considering both localization and classification tasks \cite{Lin2015}. This metric is built up three metrics of \textbf{IoU}, \textbf{precision} and \textbf{recall}.

\begin{figure}[tb]
 \centering % avoid the use of \begin{center}...\end{center} and use \centering instead (more compact)
 \includegraphics[width=\columnwidth]{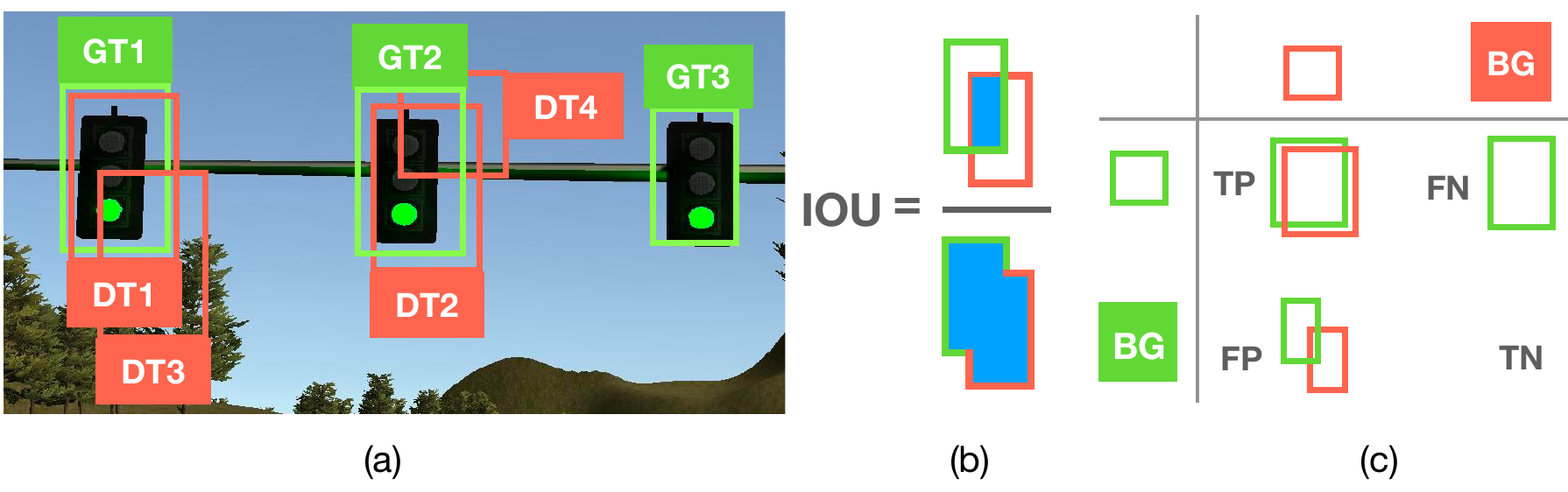}
 \vspace{-2em}
 \caption{An example of detector accuracy calculation. (a) A example with green boxes as ground truth (\textit{GT}), and red boxes as detections (\textit{DT}); (b) \textit{IoU} calculation and (c) the confusion matrix with a \textit{IoU} threshold;}
 \label{fig:detector_metrics}
\end{figure}

\textbf{IoU} (Intersection over Union) measures the overlapping between a ground truth box (\textit{GT}, green) and a detection box (\textit{DT}, red), as shown in \autoref{fig:detector_metrics}-b. By setting a threshold of IoU (usually 0.5), we can decide if a detection is a \textit{True Positive} ($TP$ with $IoU \ge 0.5$, i.e., correct detection) or a \textit{False Positive} ($FP$ with $IoU<0.5$, i.e., wrong detection), (\autoref{fig:detector_metrics}-c). If a ground truth does not have any detection, this is a \textit{False Negative} ($FN$, i.e., missed detection). In detection application, \textit{True Negatives}, namely no detection for background, are usually irrelevant for evaluation. \autoref{fig:detector_metrics}-a shows an example. The given image has three ground truth boxes ($GT_i$) and a detector yields four detections ($DT_j$) with confidence scores larger than $0.5$. With a IoU threshold of 0.5, we can know $DT_1$ and $DT_2$ are $TP$s, and $DT_3$ and $DT_4$ are $FP$s. $GT_3$ has never been detected, namely $FN$. 

%The detection results are usually sorted by confidence sores and then their IoUs can be calculated. 
% By using a threshold of 0.5, we have all positive and negative prediction, and then can compute precision and recall (\autoref{fig:detector_metrics}(c)). 

% Finally, \textbf{Average Precision (AP)} is calculated by the area under the precision-recall curve (\autoref{fig:detector_metrics}(d)). For an IoU threshold of 0.5, it is written as \textit{AP@50}. For different categories of objects (red, green, yellow), AP can be calculated as \textit{$AP_{red}$, $AP_{green}$, $AP_{yellow}$}. If we average all AP values over different IoU thresholds or categories, we can obtain a \textit{mAP} (mean Average Precision). 
Then, we can compute precision and recall: $precision=TP/(TP+FP)$, $recall=TP/(TP+FN)$. Finally, \textbf{Average Precision (AP)} is defined by the area under the precision-recall curve. For an IoU threshold of 0.5, it is written as \textit{AP@50}. If averaging all \textit{AP} values over different IoU thresholds or categories, we can obtain a \textit{mAP} (mean \textit{AP}). However, in practice, it is by far not enough to have a single aggregated metric, such as \textit{AP} and \textit{mAP}, to assess, understand and improve a detector. 

\subsection{Design Requirements}
We have outlined two challenges on both accuracy and robustness to assess, understand and improve traffic light detectors. Targeting the two challenges, we distill two categories of design requirements for our visual analytic approach from the aspects of data and performance during several design iterations (see \autoref{tab:challenges}). 

On the data side, for both existing data and unseen data, we need: 
\vspace{-0.6em}
\begin{itemize}
\itemsep-0.3em 
\item \textit{human-friendly data representation and summarization} (\inlinep{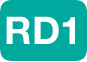}). This issue arises from the nature of high dimensionality and sheer volume of images. We need a representation to capture the intrinsic attributes of images in a lower dimension space and then summarize them in a human-friendly way \cite{LiuCNNVis, Liu2017}.
\item \textit{efficient generation} of unseen test cases (\inlinep{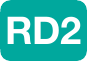}). We seek for a method generating edge cases to probe model robustness. These test cases should be different from the imperceivable noises that learned from traditional adversarial approaches, and have semantic meanings to guide human to improve the robustness \cite{Kurakin2017, Chen2019}. 
\end{itemize}

\vspace{-0.6em}
For the performance of both accuracy and robustness, we require:   
\vspace{-0.6em}

\begin{itemize}
\itemsep-0.3em 
\item a \textit{contextualized understanding} for model performance (\inlinep{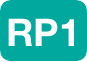}). Instead of using an aggregated metric to evaluate models \cite{Hoiem}, we would like to put a single score into the contexts of various sizes, \textit{IoU} thresholds and confident score ranges. 

\item \textit{performance interpretation} for both accuracy and robustness (\inlinep{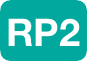}). It is challenging to understand how the accuracy and robustness of detectors are impacted by different semantic characteristics of data \cite{Zhang2018, Liu2017, Choo2018}, including colors of traffic lights (red, green, yellow, off), illumination settings (sun glare or darkness in the tree), distances (large or small size object in the scene), or confusing background (similar but irrelevant objects). 

\item \revisionblue{\textit{injecting human intelligence} for performance improvement with \textit{minimal human interaction} (\inlinep{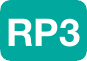}). The ultimate goal of model evaluation and interpretation is to improve its performance with human knowledge in the loop \cite{Endert2014}. Meanwhile, as we are working with domain experts, we found they have limited bandwidth to conduct in-depth exploration, but focus on key insights with few interactions \cite{Carroll1990}. This calls for maximazing insight generation and injection with minimal interaction. 
%The work towards this end for detectors demands more investigation.
} 

\end{itemize}
\vspace{-0.6em}

The two aspects are inherently related: accuracy understanding relies on existing data (training/validation/testing data), and robustness evaluation counts on unseen data. In this work, we strive for a unified visual analytic approach to alleviate these issues. 

\begin{table}[tb]
  \caption{Design requirements for \systemname\ }
  \label{tab:challenges}
  \includegraphics[width=\columnwidth]{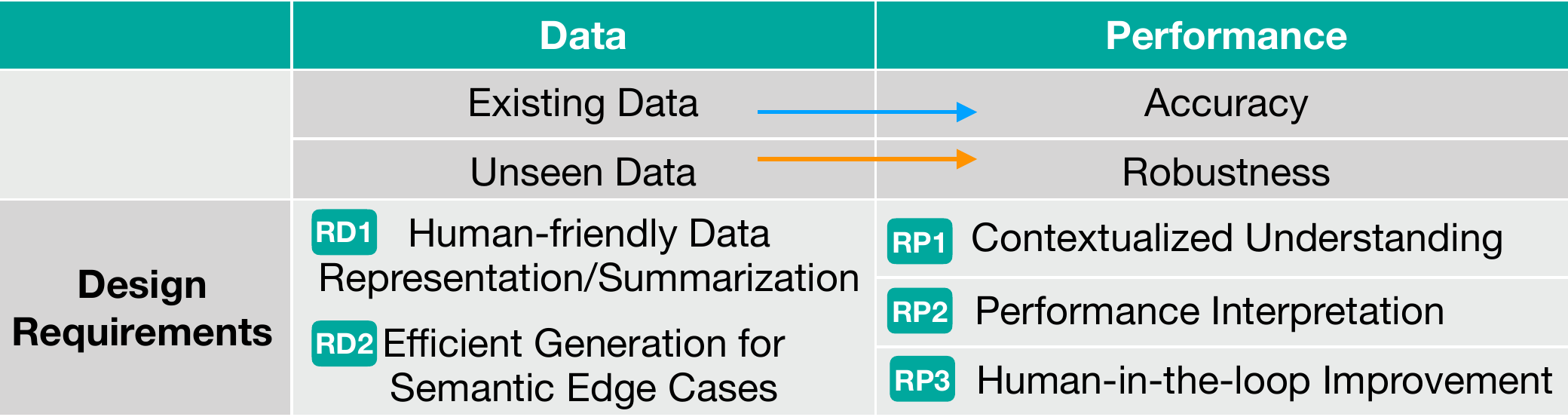}
  \vspace{-2em}
\end{table}

\section{The Framework of \systemname}
\subsection{Framework Overview}

\begin{figure*}[tb]
    \centering 
    \vspace{-2em}
    \includegraphics[width=1.9\columnwidth]{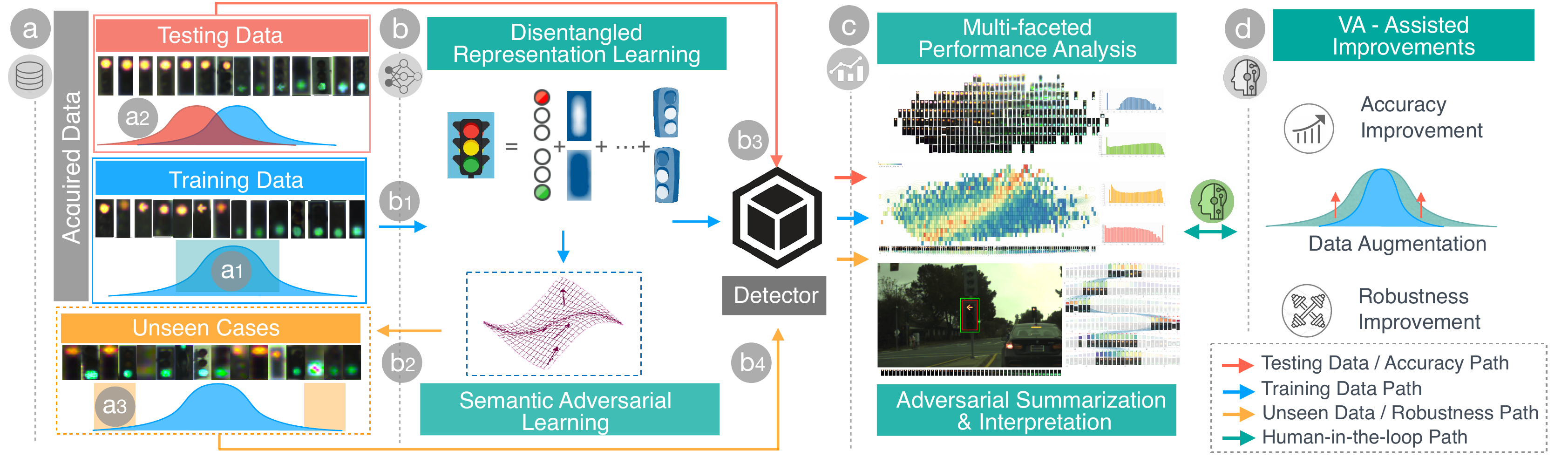}
    \vspace{-0.6em}
    \caption{The \systemname\ framework. \revisionblue{(a). Traffic light objects are first extracted from \textbf{acquired data} including both ``in distribution" training data (a1) and testing data with possible ``\textit{shifted distribution}" (a2), and also \textbf{unseen cases}, ``\textit{out of distribution}" (a3), need to be generated and tested. (b). \textbf{Disentangled representation learning} extracts interpretable attributes (e.g. colors, brightness, and background) from training objects (b1); and \textbf{semantic adversarial learning} probes the prediction behaviors (i.e. gradients) of a black-box like detector, and generates meaningful adversarial unseen examples (b2). Both \textit{acquired data} (b3) and \textit{unseen data} (b4) are passed to the detector for detection results. (c). The results along with meta-data (object size, disentangled representation, gradients etc) are transformed into interactive visualization with \textbf{multi-faceted performance analysis} and \textbf{adversarial summarization and interpretation}; (d) With minimal human interaction, actionable insights are derived to ``\textit{lift distribution up}" via \textbf{data augmentation} and inject human intelligence to improve accuracy and robustness.}}
    \vspace{-1em}
    \label{fig:framework}
\end{figure*}

We propose a framework, shown in \autoref{fig:framework}, to support domain experts assess, understand and improve the performance of traffic light detectors.  The framework consists of four modules, including \circled[0.6]{a} data processing, \circled[0.6]{b} data representation and adversarial learning, \circled[0.6]{c} interactive visualization, and \circled[0.6]{d} VA (visual analytics)-assisted improvements. 

The framework starts with \textbf{acquired data}, including training and testing data. Suppose training data is large enough to cover most cases of traffic lights a detector may encounter, such as different colors (red, green, yellow), brightness, and symbols (circle or arrows). Let's consider these cases as ``\textit{in-distribution}", as shown in \autoref{fig:framework}-\circledtwo[0.6]{a}{1} (Here, we only focus on the objects of traffic lights, not the whole scene images, because most detectors, as well as \textit{SSD}, only look for local features to locate and recognize objects). However, in testing data, there may be some traffic lights with ``\textit{shifted distribution}" causing false or missing detection, such as brightness variations, shown in \autoref{fig:framework}-\circledtwo[0.6]{a}{2}. In this work, we need to identify and understand these distribution shifts to diagnose detectors. 

Meanwhile, there are still some \textbf{unseen cases} coming ``\textit{out of distribution}" (\autoref{fig:framework}-\circledtwo[0.6]{a}{3}) that may fail a detector, and we need to generate and test them efficiently. These unseen cases may be associated with various factors, such as lighting, camera condition, and weather changes. %The generation, evaluation and interpretation for unseen cases are one of the main foci of this work, and are addressed in following modules. %We need to assess the potential risks of these cases over the detector, namely, the robustness of detector.  

In the second module, two core learning components, \textbf{disentangled representation learning} and \textbf{semantic adversarial learning}, are introduced to augment our analysis and meet the requirements of \inlinep{figures/cd1.pdf} and \inlinep{figures/cd2.pdf}. \textit{Disentangled representation learning} extracts intrinsic and interpretable attributes of traffic lights, such as colors, brightness, and background (\autoref{fig:framework}-\circledtwo[0.6]{b}{1}). This component first provides a human-friendly data presentation, and also offers a data space where an adversarial generation can efficiently search. \textit{Semantic adversarial learning} learns prediction behaviors of a detector, and generates meaningful adversarial examples (\autoref{fig:framework}-\circledtwo[0.6]{b}{2}) on top of \textbf{disentangled representation learning}. After this, both \textit{acquired data} (\autoref{fig:framework}-\circledtwo[0.6]{b}{3}) and \textit{unseen data} (\autoref{fig:framework}-\circledtwo[0.6]{b}{4}) are passed to the detector to obtain detection results. 

The third module transforms detection results, as well as the meta-data (object size, disentangled representation, gradients etc), into interactive and human-friendly visualizations (\autoref{fig:framework} -3a), including \textbf{multi-faceted performance analysis} and \textbf{adversarial summarization and interpretation}. The visualizations are designed to address performance analysis and interpretation requirements (\inlinep{figures/cp1.pdf} and \inlinep{figures/cp2.pdf}) for both accuracy over acquired data and robustness over unseen data.  

\revisionblue{Finally, with minimal human interaction from visual interface, actionable insights are derived to generate more data that attempt to ``\textit{lift distribution up}" via \textbf{data augmentation} (\autoref{fig:framework}-\circled[0.6]{d}). This also enables us inject human intelligence to improve model accuracy and robustness, aiming at the requirement \inlinep{figures/cp3.pdf}.}

% \begin{figure*}[tb]
%  \centering 
%  \vspace{-1em}
%  \includegraphics[width=1.9\columnwidth]{figures/framework.pdf}
%  \vspace{-0.6em}
%  \caption{\systemname\  framework. (a). Traffic light objects are first extracted from \textbf{acquired data} including both training data ``in distribution (a1)" and testing data with possible ``\textit{shifted distribution}"(a2), and also \textbf{unseen cases}, ``\textit{out of distribution}"(a3), need to be generated and tested. (b). \textbf{Disentangled representation learning} extracts the intrinsic and interpretable attributes (e.g. colors, brightness, and background) from the training objects (b1); and \textbf{semantic adversarial learning} obtains the prediction behaviors (i.e. gradients) of a black-box like detector, and generates meaningful adversarial unseen examples (b2). Both \textit{acquired data} (b3) and \textit{unseen data} (b4) are passed to the detector for detection results. (c). The results along with meta-data (object size, disentangled representation, gradients etc) are transformed into interactive visualization with \textbf{multi-faceted performance analysis} and \textbf{adversarial summarization and interpretation}; (d) With minimal human interaction, actionable insights are derived to ``\textit{lift distribution up}" for \textbf{data augmentation} (d1) and inject human intelligence to improve accuracy (d2) and robustness (d3).}
%  \vspace{-1em}
%  \label{fig:framework}
% \end{figure*}

\begin{figure}[tb]
    \centering 
    \vspace{-1em}
    \includegraphics[width=0.9\columnwidth]{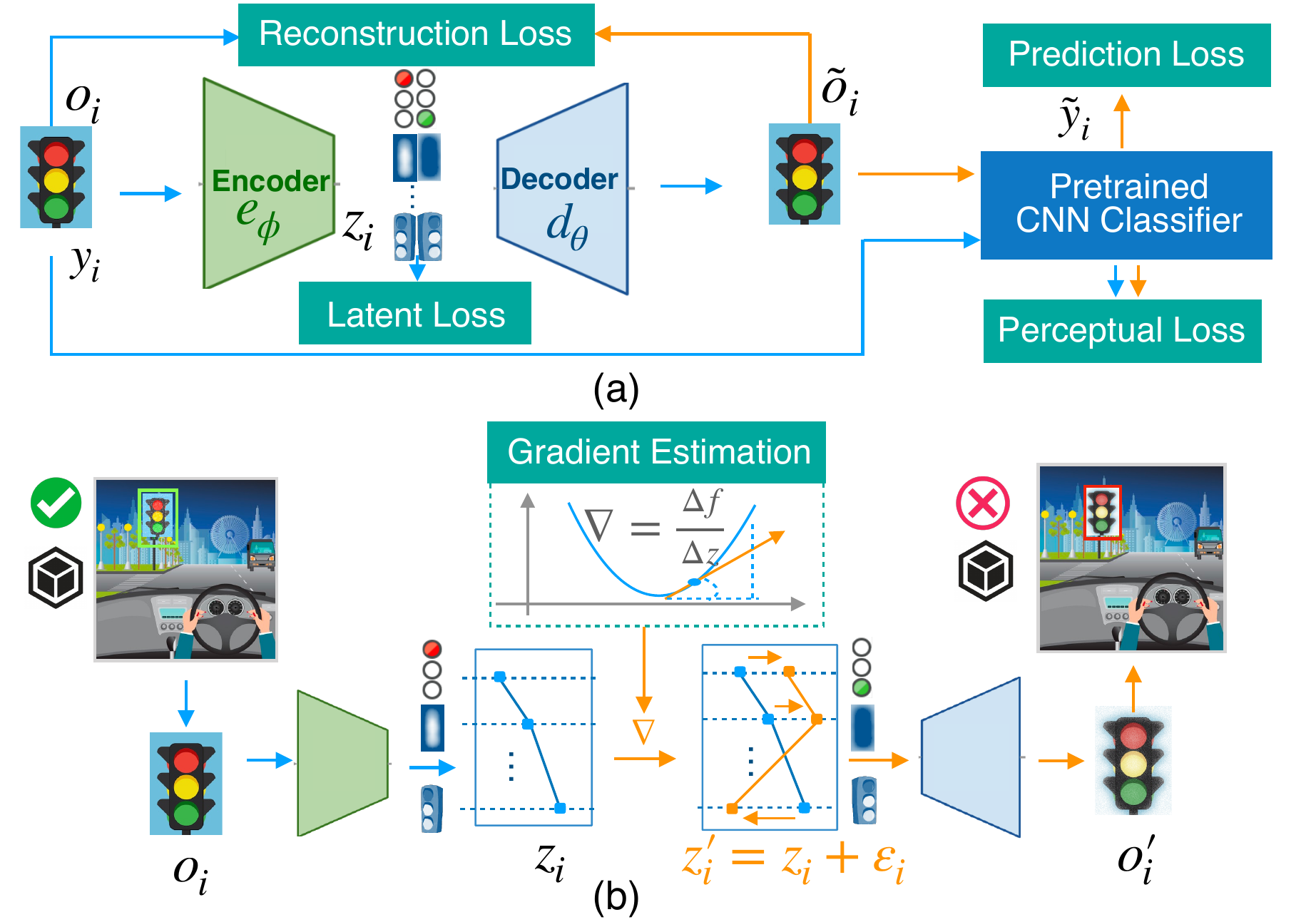}
    \vspace{-1em}
    \revisionblue{\caption{(a) Disentangled representation learning; (b) Semantic adversarial learning.}}
    \label{fig:models}
    \vspace{-2em}
   \end{figure}

\subsection{Disentangled Representation Learning (\textit{DRL})} \label{sec_drl}
Disentangled Representation Learning (\textit{DRL}) is introduced to extract semantic latent representation of traffic lights (e.g. colors, background, rotation, etc.), shown in Fig.5-a 
%\autoref{fig:models}-a, 
and generate more data with controllable semantics for data augmentation. The semantic latent representation also serves as a cornerstone for both human-friendly data summarization (\inlinep{figures/cd1.pdf}) and semantic adversarial learning (\inlinep{figures/cd2.pdf}). 

We adapted a state-of-the-art ${\beta}{\mhyphen}VAE$ \cite{Burgess2018} with customized regularization of losses. Given an object image ($o_i \in O^N$, a traffic light with size $N{\times}N$), ${\beta}{\mhyphen}VAE$ includes two components: an encoder, $e_{\phi}$, mapping an input ($o_i$) into a latent vector ($z_i \in Z^D$, $D$: latent dimension size), namely $e_{\phi} \colon o_i \mapsto z_i $ , and a decoder ($d_{\theta}$) converting a latent vector ($z_i$) into a reconstructed image ($\tilde{o}_i$), namely $d_{\theta} \colon z_i \mapsto \tilde{o}_i $. %(Strictly, a latent vector ($z_i$) consists of a mean vector ($\mu_i$) and a standard deviation vector ($\sigma_i$):  $z_i = \mu_i +\sigma_i $. In this work, we safely ignore $\sigma_i$, and use $z_i$ and $\mu_i$ equivalently, namely $z_i \sim \mu_i$.)  

In the vanilla ${\beta}{\mhyphen}VAE$, it uses two losses to optimize the model including \textbf{reconstruction loss} (usually, a mean square error: $MSE = {\|o_i - \tilde{o}_i\|^2}$) and \textbf{latent loss} (KL divergence: $D_{KL} =D_{KL}((z|x)\|(z))$). 

\revisionblue{Here, we introduce two more regularization terms of \textbf{prediction loss} and \textbf{perceptual loss} \cite{Hou} to generate realistic images. 
%The idea is to improve both reconstruction quality and the semantic expressiveness of the latent space with additional information, such as the label of traffic lights. 
We first train a CNN based classifier to predict traffic light colors.  
%(currently using light status of red, green, yellow and off, but they can be any available labels).
Then, we employ the pre-trained CNN classifier to predict the color ($\tilde{y}_i$) of a reconstructed image $\tilde{o}_i$, and also extract the feature maps from ConvNet layers ($\phi^l(x)$ from the $l$th ConvNet layer) of both original ($o_i$) and reconstructed images ($\tilde{o}_i$). Finally, we can obtain \textit{prediction loss} (Cross-Entropy loss: $CE(y_i, \tilde{y}_i)$) and \textit{perceptual loss} \cite{Hou} of feature maps between original and reconstructed images ($\Sigma^L {\|\phi^l(o_i) - \phi^l(\tilde{o_i})\|^2}$). }

The final loss term is the sum of all losses introduced above: $\mathcal{L} = MSE(o_i, \tilde{o}_i) + \gamma |D_{KL} - C| + \mu CE(y_i, \tilde{y}_i) + \nu  \Sigma^L {\|\phi^l(o_i) - \phi^l(\tilde{o_i})\|^2}$, where $\gamma, C$ are parameters to control disentanglement, and $\mu, \nu$ are weights to control reconstruction quality. 
%With additional regularization, the quality of reconstruction and latent distribution is improved (see supplemental materials).

\subsection{Semantic Adversarial Learning}
Built upon \textit{DRL}, we propose a novel semantic adversarial learning (\textit{SemAdv}) method to generate meaningful adversarial examples to test and interpret the robustness of a detector (Fig.5-b). 
%\autoref{fig:models}-b. 
Two unique requirements should be met: a). Adversarial examples need to be meaningful (not imperceivable noise) and can guide us improve model robustness; b). The method should be applied to any detector without accessing model parameters, namely a model-agnostic approach. 

% The idea is that, given a traffic light object ($o_i$) from a driving scene ($x_j$) and a black-box detector ($f$), \textit{SemAdv} needs to search the latent space, $Z^D$, to find a $z_i'=z_i+\varepsilon_{i}$ with the minimal change (the smallest $\varepsilon_{i}$) that can generate an adversarial object image, $o'_i = d_{\theta}(z_i')$ to successfully fail the detector, namely, $f(o'_{i\rightarrow} x_j)<0.5$. 
\revisionblue{The idea is that, given a traffic light object ($o_i$) from a driving scene ($x_j$) and a black-box detector ($f$), \textit{SemAdv} needs to generate an adversarial object image, $o'_i$, to fool the detector, namely, $f(o'_{i\rightarrow} x_j)<0.5$. Here, $f(o'_{i\rightarrow} x_j)$ is the detection score of traffic light $o'_i$ re-inserted in the original scene $x_j$. For detection results, $P$, with bounding boxes, $\{b\}_p$, and confidence scores, $\{c\}_p$, the $f(o'_{i\rightarrow} x_j)$ is defined as:}
\vspace{-0.6em}
\begin{equation}
f(o'_{i\rightarrow} x_j) = \begin{cases}
max\{c_d\} &\mbox{if } IoU(b_i,b_d) \geq 0.5, \forall d \in P \\
0 &\mbox{if } IoU(b_i,b_d)<0.5, \forall d \in P
\end{cases}
\end{equation}
\vspace{-1em}

\revisionblue{The core step is to searching the object's latent space, $Z^D$, to find a $z_i'=z_i+\varepsilon_{i}$ with a minimal change to generate the adversarial $o'_i$. The algorithm is shown in \autoref{algorithm:SemAdv}.}

\revisionblue{We adapt a Fast Gradient Sign Attack (FGSM) based approach \cite{Goodfellow2015} to obtain the minimal shift $\varepsilon$ (\textit{Line 8-15} in \autoref{algorithm:SemAdv}). (FGSM is computationally efficient compared to other strong attack methods, e.g. PGD \cite{Madry2018}, by considering the high cost of gradient estimation needed.) The main idea is to obtain the gradient of a detector, $\nabla$, and then iteratively push the latent vector, $z_i$, to move along the ascending direction of the gradient, $ z_i^{t+1} = z_i^{t} - \nabla $, until the reconstructed object fails the detector in the scene by $f(o'_{i\rightarrow} x_j)<0.5$.} %This process is exactly opposite to optimize a model with descending gradient during training.

\revisionblue{To obtain the gradient without knowing detector parameters, we use a black-box gradient estimation approach by Natural Evolution Strategies (NES) \cite{Wierstra2014, Li-NATTACK} (\textit{Line 2-7} in \autoref{algorithm:SemAdv}). Here, the gradient is approximated by the amount of output changes, $\Delta f$, caused by input changes, $\Delta z$, namely, $\nabla={\Delta f}/{\Delta z}$. Finally, we constraint the change for each latent dimension smaller than a budget $\lambda$, namely $\varepsilon_{i,d}<\lambda, \forall d \in Z^D$, to make sure a reconstructed image not goes unrealistic. }
% This budget is usually defined by the scale of the dimension standard deviation. 

%In NES, we perturb the latent vector, $z_i+\{\delta\}_k$, with $k$ noise samples, and then obtain the changes of detector scores, $\Delta f$. With a similar fashion of calculating the slope of the tangent line passing point ($z_i, f_i$), we can estimate the gradient at this data point, $\nabla={\Delta f}/{\Delta z}$, where $\Delta z \sim \{\delta\}_k$. 

\vspace{-0.5em}
\newcommand\mycommfont[1]{\footnotesize\ttfamily\textcolor{blue}{#1}}
\SetCommentSty{mycommfont}

\SetKwInput{KwInput}{Input}                % Set the Input
\SetKwInput{KwOutput}{Output}              % set the Output

\begin{algorithm}[h]
    \DontPrintSemicolon
    
    \KwInput{an object $o_i$, from a scene $x_j$, a detector $f(x)$, an encoder $e_{\phi}$, a decoder $d_{\theta}$, learning rate $\eta$, perturbation sample size $K$, perturbation scale $\delta$, maximal iteration steps $T$, and maximal modification budget $\lambda$ }
    
    \KwOutput{an adversarial object $o'_i$, and its modification  $\varepsilon_{i}$ }
    
    \If{$f({o}_{i\rightarrow} x_j)<0.5$}
        {
            \Return ``Already failed. No need to attack!''
        }
    Encode object $ z_i = e_{\phi}(o_i)$ 
 
    \tcp{Estimate gradient for $ z_i$}
    Sample $\sigma_1, ... , \sigma_k \sim \mathcal{N}(0,I)$ for $K$ times
    
    Get perturbation:  $ z_{ik} = z_i + \delta \sigma_k, \forall k$
    
    Generate perturbed objects:  $ \tilde{o}_{ik} = d_{\theta}(z_{ik}), \forall k$
    
    Detection scores: $ f_k =  f(\tilde{o}_{ik\rightarrow} x_j), \forall k$
    
    Normalized scores: $ \hat{f_k} = (f_k-mean(f))/std(f), \forall k$
    
    Gradient estimation: $\nabla =\frac{1}{K\delta} \Sigma^{K}_{k=1}{\sigma_k \hat{f_k}}$
    
    \tcp{Update $z_i^{}$ by $T$ steps}
    \For{$t\gets0$ \KwTo $T$}{ 
        
        $ z_i^{t+1} = z_i^{t} - \eta \nabla $ 
        
        Constrain by budget: $z_i^{t+1} = clip(z_i^{t+1}, \lambda)$
        
        Generate a new object: $ \tilde{o}_i^{t+1} = d_{\theta}(z_i^{t+1})$
        
        \If{$f(\tilde{o}^{t+1}_{i\rightarrow} x_j)<0.5$}
        { 
            \tcp{found an adversarial $o'_i$}
            $ o'_i = \tilde{o}^{t+1}_i$ 
            
            Compute the change: $ \varepsilon_{i} = z_i^{t+1} - z_i^{0} $ 
            
            \Return $o'_i$, $\varepsilon_{i}$
        }
    }
    \tcp{Failed within $T$ steps and modification budget $\lambda$}
    
    Set the change as the max budget: $\varepsilon_{i} = \lambda$
    
    \Return  $o'_i$, $\varepsilon_{i}$
    
\caption{\revisionblue{Semantic Adversarial Learning (\textit{SemAdv})}}
\label{algorithm:SemAdv}

\end{algorithm}
\vspace{-1em}

\textbf{\textit{Semantic Robustness}}.
With \textit{SemAdv} method, we can define a semantic robustness score of a detector by measuring how much minimal change in the latent space we need to make to fail the detector. For a specific object, $o_i$, the semantic robustness is the average L1-norm of normalized minimal change in the latent space: $sr_i = \frac{\| \varepsilon_{i}/{std(Z)} \|_{L1} }{\lambda D}$, where ${\varepsilon_{i}}/{std(z)}$ means that each dimension of $\varepsilon_{i}$ is normalized by the standard deviation of each dimension, $\|*\|_{L1}$ is the \textit{L1-norm}, $\lambda$ is the modification budget and $D$ is the dimension size. The semantic robustness of a detector, $sr(f)$, is defined as the average semantic robustness for all objects, $O$, in a dataset: $sr(f) =\Sigma^{O}_{i=1}{sr(o_i)}/|O|, \forall i \in O$.

% \begin{equation}
% sr(f) = \frac{1}{|O|} \Sigma^{O}_{i=1}{\|\frac{\varepsilon_{i}}{std(z)}\|_{L1}}, \forall i \in O
% \end{equation}
% , where ${\varepsilon_{i}}/{std(z)}$ means that each dimension of $\varepsilon_{i}$ is normalized by the standard deviation of each dimension, $\|*\|_{L1}$ is the \textit{L1-norm}, and $D$ is the dimension size. 

This semantic robustness is different from traditional adversarial robustness because it is meaningful and interpretable. We can intuitively know how much effort we need to change on each dimension with visual semantics, such as color, lightning, background, to fail a detector.  

\subsection{Data Extraction}
%We summarize the structure of data extracted for visual analytics in the framework, shown in \autoref{tab:data_structure}. 
\revisionblue{We extract three types of data, including \textit{top-n} detections, ground truth objects, and semantic adversarial objects, for analysis. We summarize the structure of data extracted for visual analytics in \autoref{tab:data_structure}}. For each type of data, we collect performance information (bounding boxes, confidence sores, \textit{IoU}s), data representation (image patches, latent vectors) and other meta-information (size, class, gradient, robustness). For unseen adversarial data, we only learn adversarials for the corresponding ground truth objects that have been successfully detected, namely $f({o}_{i\rightarrow} x_j)>0.5$ (\textit{Line 8-15} in \autoref{algorithm:SemAdv}). 

\begin{table}[tb]
    \vspace{-0.5em}
    \revisionblue{\caption{Data extracted for visual analytics in the framework}}
  \label{tab:data_structure}
  %\vspace{-0.5em}
  \includegraphics[width=\columnwidth]{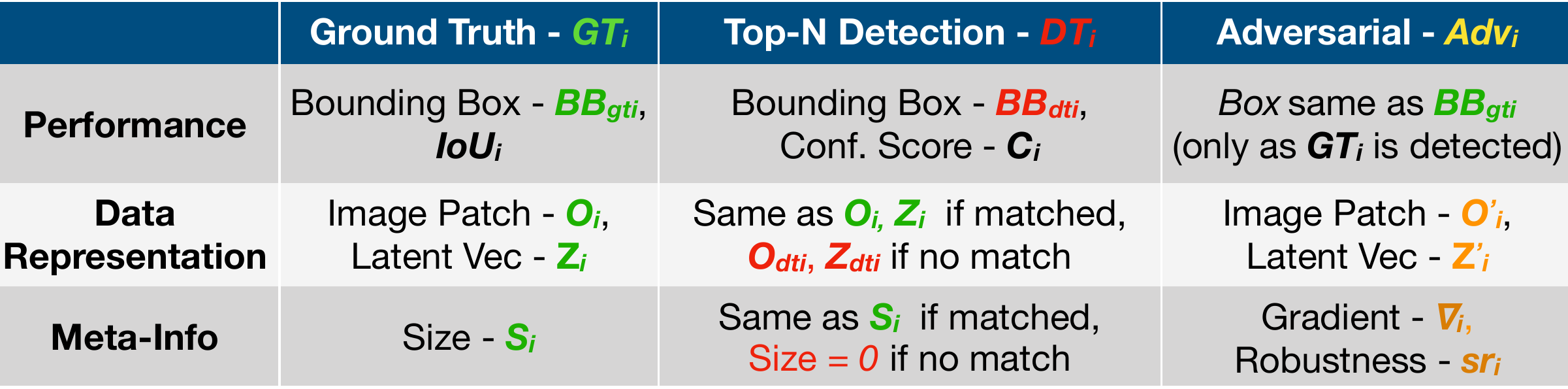}
  \vspace{-2.5em}
\end{table}

\section{The User Interface of \systemname}
Guided by above framework, we design and implement a visual analytics system, \systemname, \autoref{fig:teaser}, to assess, understand and improve traffic light detectors. The system user interface consists of four views: \circled[0.6]{a} \textit{Multi-faceted Summary}, \circled[0.6]{b} \textit{Performance Landscape-\tilescape}, \circled[0.6]{c} \textit{Driving Scene View}, and \circled[0.6]{d} \textit{Semantic Representation View-\hpcp}.

\subsection{Multi-faceted Summary}
\revisionblue{
This view offers a summary and navigation for key performance statistics ( \autoref{fig:teaser}-\circled[0.6]{a}). The top of \autoref{fig:teaser}-\circled[0.6]{a}, summarizes the total number of traffic light objects in a dataset, the number of \textit{top} detections per image (top-5 in this work), false positive (\textit{FP}), false negative (including both never detected lights, \textit{FN-I}, and confusing lights with low confidence scores, \textit{FN-II}) and total adversarial. Also, bar charts in \autoref{fig:teaser}-\circled[0.6]{a}, show the distribution of performance metrics (including IoU, confidence score, and robustness for valid detection) and meta-info (object size). }

\revisionblue{
Furthermore, the metric charts are coordinated with each other to filter data and support multi-faceted performance analysis for accuracy and robustness in other views. This is designed to support \textit{contextualized understanding} of performance, \inlinep{figures/cp1.pdf}.
} 
For example, \textit{FP}, \textit{FN-I} and \textit{FN-II} detections can be selected by setting different filtering conditions: \textit{FP}s are detections with zero size (no ground truth), \textit{IoU} larger than a threshold of 0.5, and confidence score larger than 0.5, namely, $FP =\{DT_i|s_i=0\wedge IoU_i>0.5 \wedge c_i>0.5 \}$. %Similarly, we have \textit{FN-I}$=\{DT_i|s_i>0\wedge IoU_i=0 \wedge c_i=0\}$, \textit{FN-II}$=\{DT_i|s_i>0\wedge IoU_i>0.5 \wedge c_i<0.5\}$. 
This view also enables a user to quickly navigate to \textit{FP} or \textit{FN} visual summary by clicking the icons in the dashboard to apply pre-defined filtering conditions. 
%Additionally, performance analysis for understanding the impact of object size is also supported by filtering the detection with various sizes. 
%This is also applied to other available meth-info, such as aspect ratio, distance, rotation etc. 

%The performance analysis of other factors is also supported. 

\subsection{Performance Landscape View: \tilescape}
\begin{figure}[tb]
 \centering 
 \vspace{-0em}
 \includegraphics[width=\columnwidth]{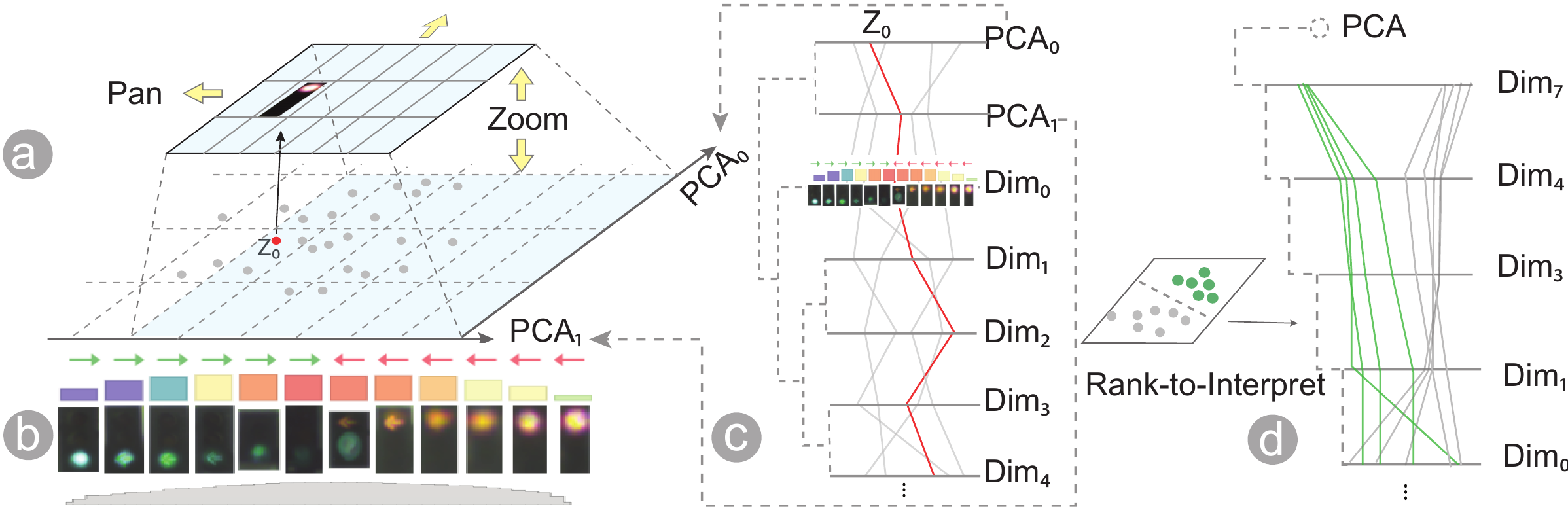}
 \caption{The design \tilescape\ and \hpcp: a) Tile aggregation; b) Dimension visual encodings; c) Hierarchical PCP; d) Rank-To-Interpret. }
 \vspace{-2em}
 \label{fig:tilescape_hpcp}
\end{figure}

We design a view of performance landscape, \tilescape, to summarize visual characteristics and corresponding performance over tens of thousands of objects \revisionblue{(design requirement \inlinep{figures/cp2.pdf})}, shown in the panel \circled[0.6]{b} of \autoref{fig:teaser}. 
In \tilescape, each cell (tile) is an aggregated bin of many detections which are similar in the semantic latent space learned from \textit{DRL} (\autoref{fig:tilescape_hpcp}-\circled[0.6]{a}). Each data-point is a detection, located in the latent space by its latent vector encoded from a detected image patch. Then, the data-points are aggregated into bins, called as \textit{tiles}, according to a view range and bin size. Also, one object is selected as the representative one for each tile. Here, we use the object with median score from a bin. Additionally, in the background of \tilescape, a contour density map shows the data distribution in the current space. 

Each tile visually encodes one of four categories of object information: \textit{visual appearance} (image patch), \textit{confidence score}, \textit{robustness score} and \textit{adversarial gradient direction}. As shown in the upper part of \autoref{fig:teaser}-\circled[0.6]{b}, each tile shows a detected image patch: either a ground truth (\textit{GT}) object image when a detection (\textit{DT}) is matched, or the image patch cropped from a driving scene by a \textit{DT} bounding box when no matched \textit{GT} object is found for the current \textit{DT}. The tile can also encode detection accuracy and robustness scores with colored rectangles, shown in the lower part of \autoref{fig:teaser}-\circled[0.6]{b}. %For adversarial gradients, we design a glyph of arrow to encode the directions \autoref{fig:teaser}-\circledtwo[0.6]{b}{3}. 
Users can change visual encoding for tiles via a drop-down menu in the panel. 

Along the two axes of \tilescape, we also design an aggregated image bars to show the data distribution along each axis, \autoref{fig:tilescape_hpcp}-\circled[0.6]{b}. The data is also binned and aggregated with the same approach described above for each axis. Each bar consists of a representative image, a rectangle and an arrow glyph with similar visual encoding: bar height indicates the number of data-points, bar color encodes the median score of either confidence or robustness, and an arrow shows gradient direction. 

Users can zoom and pan \tilescape, to observe the patterns at different levels of aggregation. Also, users can interactively select any latent dimensions or the PCA components of latent dimensions to lay out the \tilescape\ view. By default, the first two PCAs of latent dimensions are used to show overall visual and performance patterns. 

%\textbf{Adversarial Landscape}

\subsection{Semantic Representation View: \hpcp}
To efficiently explore the semantic representation space, a hierarchical parallel coordinates plot, (\hpcp), is designed as shown in \autoref{fig:teaser}-\circled[0.6]{d}. First, latent dimensions are hierarchically clustered with an agglomerative method to efficiently organize and navigate these dimensions (\autoref{fig:tilescape_hpcp}-\circled[0.6]{c}). The clusters are formed by a ward linkage \cite{Ward1963} that minimizes the variance of Euclidean distance among all latent object vectors within a cluster. Only top dimensions can be visible by applying a distance threshold. More dimensions can be shown by expanding subtrees. The first two PCA components of latent dimensions are also included to capture the dominate variance of all dimensions. They are organized as a special subtree in the root node. 
%An illustration is shown in the right of \autoref{fig:tilescape_hpcp}. 
 
An aggregated image bar plot is also used for each dimension to show the semantics captured by this dimension, shown in \autoref{fig:tilescape_hpcp}-\circled[0.6]{b}. We first bin and aggregate the data over each dimension, and then use the same visual encoding for data distribution, performance (confidence and robustness) scores and gradient directions as in \tilescape. The shown image here is a blending of five random underlying images. Additionally, a gray area plot is introduced as background to compare data distribution with the forefront bar chart. We can use this to compare data distribution of testing and training data over each dimension.

A parallel coordinates plot is then applied upon the top dimensions. By connecting data coordinates across all visible dimensions with curved lines, it shows dimension correlation and data clusters. To reduce visual clutter, we do not show the lines of all data-points but only these of our interest via users' interaction. For example, hovering a bin on a dimension will show lines for the dat points in this bin. 

\textbf{Semantic interpretation by coordinating with \tilescape}. Two mechanisms are introduced to help users understand dimension semantics and also interpret their impact over performance (\inlinep{figures/cp3.pdf}).

\textit{\hpcp}\ $\mathbf{\rightarrow}$ \textit{\tilescape}. Any dimension from \hpcp\ can be selected as $x$ or $y$ axis for \tilescape\ to examine what visual semantics are embedded in this dimension. Also by hovering or brushing the dimension bins in \hpcp,  corresponding tiles in \tilescape\ are highlighted to show what visual feature the selected bins capture. 

\textit{\tilescape}\ $\mathbf{\rightarrow}$ \textit{\hpcp}. Interaction from \tilescape\ to \hpcp\ enables users to interpret detector's performance with the semantics learned by latent dimensions. Users can use a lasso to select a group of bins of interest (e.g. low confidence scores) in \tilescape, and corresponding lines and bins are highlighted in the latent space of \hpcp. 

A \textbf{Rank-To-Interpret} method is proposed to rank dimensions by their importance to separate the selection from other data points (\autoref{fig:tilescape_hpcp}-\circled[0.6]{d}). The selected and unselected data are first marked with different labels for a target variable, and their latent vectors are used as features to estimate their mutual information (\textit{MI}) towards the target variable \cite{Ross2014}. The dimensions are then ranked by their \textit{MI} values and agglomeratively organized as a tree structure for \hpcp. In this way, we can understand the top semantic dimensions explaining the selection.

\subsection{Driving Scene View}
The driving scene view offers a live detection result for a selected object (\autoref{fig:teaser}-\circled[0.6]{c}). This view shows the ground truth box (green) and detected bounding box (red) in a driving scene image, and also scores of \textit{IOU}, confidence and robustness (\autoref{fig:teaser}-\circledtwo[0.6]{c}{1}). 

This view also helps users understand adversarial samples and associated adversarial gradients by inserting different adversarial generations into its driving scene to examine detection results, shown in the bottom of \autoref{fig:teaser}-\circledtwo[0.6]{c}{2}. The original object is in the middle (zero point) of the horizontal axis, and different samples are generated from this object by moving ascending (right) or descending (left) along the direction of adversarial gradient in the latent space. By examining generated images and their test results, we can see how an adversarial gradient changes object appearance and detection performance.

\begin{figure*}
 \centering 

 \includegraphics[width=2\columnwidth]{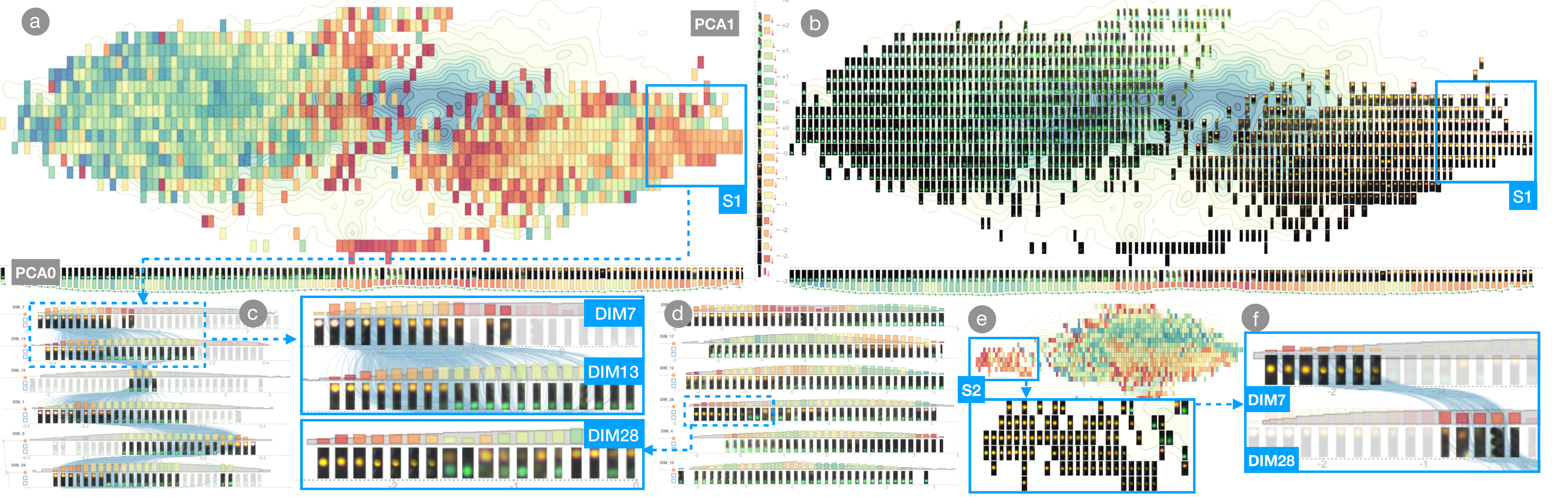}
 \vspace{-1em}
 \caption{The detection confidence (accuracy) landscape and error diagnosing over test dataset. (a) Detection confidence summary over the first two PCA components of latent dimensions, and one low confidence area (\textit{`S1'}); (b) Object visual summary over PCA components; (c) Ranked latent dimensions explain how the selection \textit{`S1'} is different from the rest, and the dimension \textit{`DIM7'}, and \textit{`DIM13'} capture distribution shift of high brightness for red lights; (d) One top ranked dimension, \textit{`DIM28'}, indicates low confidence scores over yellow lights; (e) Re-layout confidence summary with \textit{`DIM28'} as x-axis, and show an outlier cluster of yellow lights with low confidence score, \textit{`S2'}; (f) Top two ranked latent dimensions, \textit{`DIM28'} and \textit{`DIM7'}, explain the visual semantics of yellow lights with low confidence scores.}
 \vspace{-1em}
 \label{fig:test_acc_landscape}
\end{figure*}

\section{Case Study and Evaluation}
\revisionblue{We closely worked with four domain experts from a function testing department for autonomous driving applications over one year. Two of the experts collaboratively contributed this work and the other two conducted model testing in their daily work. We together went through several design iterations and system refinements to alleviate their pain-points, namely how to generate \textit{\textbf{actionable insights}} to \textit{assess}, \textit{understand} and \textit{improve} traffic light detectors with \textit{\textbf{minimal human-in-the-loop}}. The system has been successfully transferred and deployed in their testing platform. The following use cases and experimental evaluations were distilled from their routine practice of model testing.}

\subsection{The Dataset and Detector}
\revisionblue{The experts suggested to use a public dataset, Bosch Small Traffic Lights Dataset \cite{bstld_data}, to study and benchmark system capability.} The dataset includes 5093 training images (10756 annotated traffic lights) and 8334 test images (13486 annotated traffic lights). The baseline traffic light detector was provided by domain experts for case study purpose. The provided detector is based on \textit{SSD MobileNet V1} (see \autoref{fig:ssd} for model architecture) and achieves state-of-the-art accuracy with $AP@IoU50=\mathbf{0.478}$ (compared with \textit{0.41} published in \cite{bstld_repo}). We ran the detector over both training and testing dataset and saved detection results for analysis.
%\footnote{https://hci.iwr.uni-heidelberg.de/node/6132}
%\footnote{https://github.com/bosch-ros-pkg/bstld} 
%(Running detection over the training data-set is for model diagnosing and debugging). 

\subsection{Representation and Adversarial Learning}
We cropped 10683 valid traffic lights from the \textit{training} dataset and then resized them into $64$x$64$ to train the models of disentangled representation (\textit{DRL}) and adversarial learning (\textit{SeADV}). We only used training dataset for both models and reserved the test dataset for experiments. 

For \textit{DRL}, the network architecture is the same as ${\beta}{\mhyphen}VAE$ \cite{Burgess2018} with latent dimension $|Z|=32$, $\gamma=100, C=20, \mu=4, \nu=1$, learning rate 5e-5, 2K maximum iterations to increase capacity $C$, and 5k total iterations. %The CNN classifier has the same structure as the \textit{DRL} encoder with fully connected layers and a soft-max layer as output, and achieved $97.36\%$ test accuracy to predict the traffic light colors. 
With the pre-trained \textit{DRL} model, we conducted semantic adversarial attack and generation against the base detector over all \textbf{\textit{train}} objects. %As shown in \autoref{tab:model_comparison}, 
Among 7286 successful detected objects by the base detector, we obtained 6830 adversarial objects ($93.7\%$ success rate) and failed to attack 456 objects with the modification budget of two standard deviations over latent vectors, $\lambda=2$ (other hyper-parameters in \textit{SeADV}: $\eta = 0.01$, $K=512$, $\delta=0.5$, and $T=500$). 

% A PCA projection model was generated from the latent vectors of training data, and the first two PCA components were extracted for initial layout of training objects. The PCA components of testing latent vectors were also projected via this PCA model to keep the PCA views of training and testing data consistent and comparable. 
% 

%learning rate $\eta = 0.01$, perturbation sample size $K=512$, perturbation scale $\delta=0.5$ and maximal iteration steps $T=500$
%This ended with a semantic robustness of $0.499$ for the base detector. 

%We use following parameters setup for semantic adversarial learning: 

After model pretraining and data preprocessing, we extracted required data from both training and testing datasets to drive interactive visualization. The first two PCA components of both training and testing objects were projected with the same PCA model that was built upon the latent vectors of training objects. This kept the PCA views of training and testing data consistent and comparable.

% \begin{figure*}
%  \centering 
%  \includegraphics[width=2.1\columnwidth]{figures/case-rbt.pdf}
%  \caption{The semantic robustness landscape over the training dataset. (1) Detection robustness summary over two PCA components and a selection of low robustness area (\textit{`S3'}); (2) The corresponding object visual summary over two PCA components; (3) Ranked latent dimensions to explain low robustness selection \textit{`S3'} and the dimension `2' and `13' capture the semantics of dark and red lights;(4) The gradient directions shows the weakness directions pointing towards the darker areas; (5) A live test view shows the detection scores for the generated objects as the generator move upwards (+) or downwards (-) along the gradient direction.}
%  \label{fig:robustness_landscape}
% \end{figure*}

\subsection{Visual Analysis Cases}

\subsubsection{Show me performance landscape}
\vspace{-0.25em}
The experts were eager to see the overall accuracy landscape of the detector. They first examined the confidence landscape over test dataset, shown in \autoref{fig:test_acc_landscape}. Intuitively, they had several observations.

First, there was a clear distribution imbalance of detection accuracy over different object colors, as shown in \autoref{fig:test_acc_landscape}-\circled[0.6]{a}\circled[0.6]{b}. The semantics of color were captured by the first PCA component for latent dimensions, and shown as the x-axis in \autoref{fig:test_acc_landscape}-\circled[0.6]{b}: right for red and left for green light (the color variance was learned from the data, and we didn't explicitly use color attributes). It was clear that the detector had worse performance over red lights compared with green ones. 

Secondly, data distribution gap and low confidence score areas were observed. There was a large sparse area in the middle of \autoref{fig:test_acc_landscape}-\circled[0.6]{b}, indicating testing data has less coverage in this area (compared with the background density map built from the training dataset). However, one interesting finding is that the detector still had low confidence score in the middle part even though the training data had dense coverage in this part. Also, another low score area is located on very right side, \inlinep{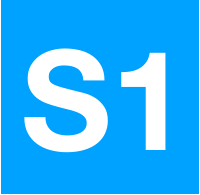} in \autoref{fig:test_acc_landscape}-\circled[0.6]{a}. This area are mostly red lights that looked clear and bright, and it was quite interesting to see why the detector failed here.  

Therefore, two following-up questions were asked by the domain experts: a) What factors lead the low performance areas? b) What actionable insights we can have to improve the accuracy of our interest?

\begin{figure}[tb]
 \centering 
 \vspace{-1em}
 \includegraphics[width=0.9\columnwidth]{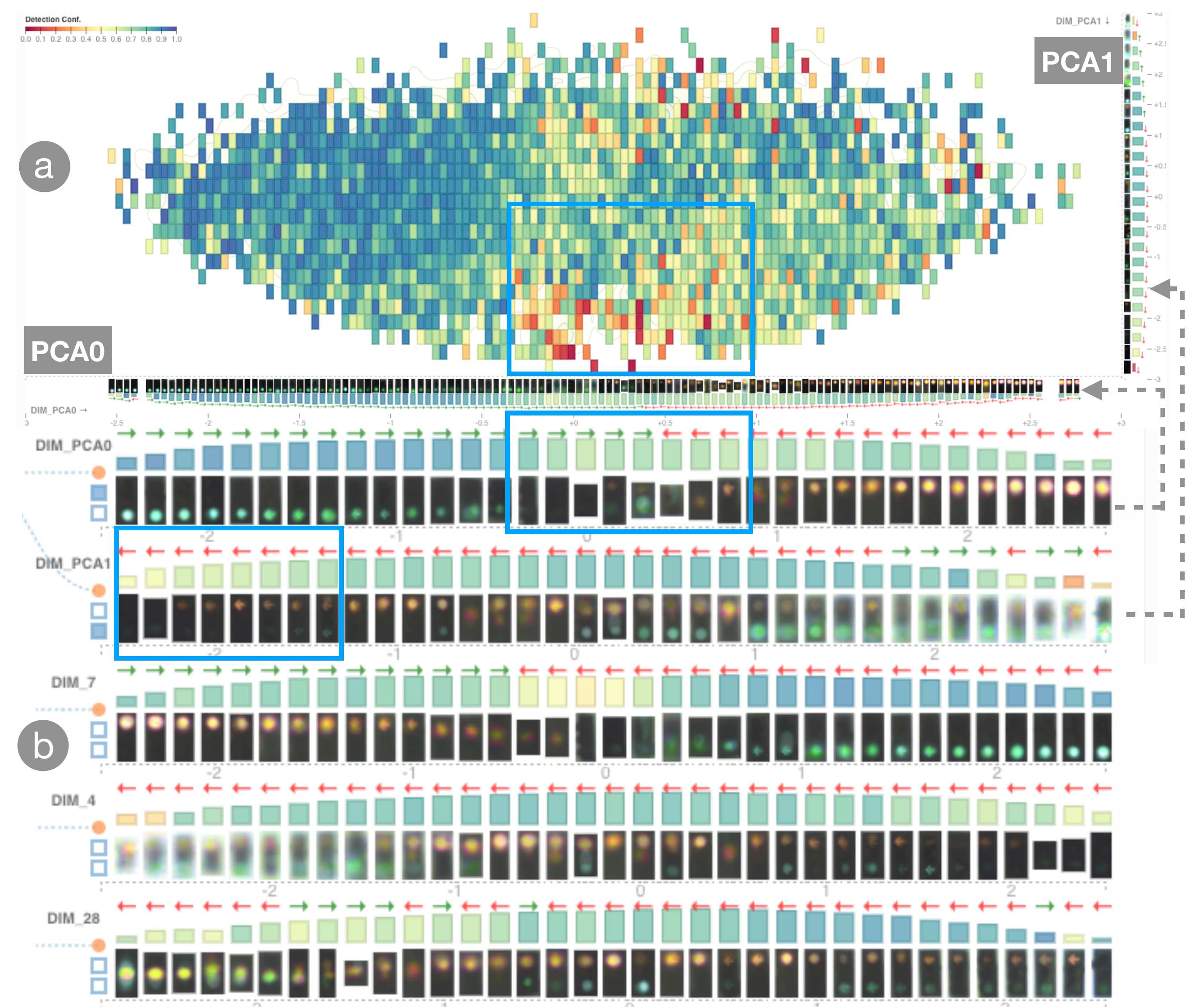}
 \vspace{-1em}
 \caption{The visual summary for training data. (a) Detection confidence (accuracy) landscape (b) Top latent dimensions with semantics.}
 \vspace{-1.5em}
 \label{fig:train_landscape}
\end{figure}

\vspace{-0.5em}
\subsubsection{What confused the detector?}
\vspace{-0.25em}
\label{case_insight_acc}
The experts were first interested in why the detector had low confidence score in the middle area where the training data already had good coverage. They turned to the training data by examining visual summary of objects (upper part in \autoref{fig:teaser}-\circled[0.6]{b}) and confidence scores (\autoref{fig:train_landscape}-\circled[0.6]{a}). They observed that even for the training data, traffic light objects in the middle bottom part was quite challenging for the detector, as shown in yellow and red score bins. By looking and exploring the semantics of top dimensions in \autoref{fig:train_landscape}-\circled[0.6]{b}, they found the middle range of \textit{PCA0} explains the ambiguity of traffic lights caused low confidence scores (as \textit{PCA0} mainly captures the variations of colors), and the lower end of \textit{PCA1} shows dark traffic lights are difficult for the detector (as \textit{PCA1} mainly explains the variations from dark to bright). Therefore, they could conclude that the ambiguous and dark traffic lights explained the low confidence area in the middle part. 

To investigate the low confidence area \inlinep{figures/s1.pdf} in \autoref{fig:test_acc_landscape}-\circled[0.6]{a}\circled[0.6]{b}, they ranked latent dimensions to show their importance contributing to this selection. As shown in \autoref{fig:test_acc_landscape}-\circled[0.6]{c}, they observed that the left end of the  \textit{`DIM7'}, and \textit{`DIM13'} with red color contributed most to the selection \inlinep{figures/s1.pdf}. They also noticed that the selection on these two dimensions has very high brightness, and the distribution of high brightness in testing data (the colored bars in \autoref{fig:test_acc_landscape}-\circled[0.6]{c}) is much larger than the distribution in training data (the gray area plot in \autoref{fig:test_acc_landscape}-\circled[0.6]{c}. This explained the low confidence scores of \inlinep{figures/s1.pdf} resulted from the under-representative high brightness red lights in the training dataset. 

Further, they also observed one dimension, \textit{`DIM28'} among the top ranked dimensions, showed low confidence scores over yellow lights (\autoref{fig:test_acc_landscape}-\circled[0.6]{d}). After using \textit{`DIM28'} as x-axis to re-lay out the score summary, they can see an outlier cluster of yellow lights with low confidence scores, shown as \inlinep{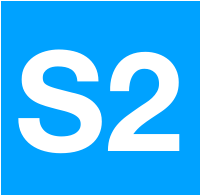} in \autoref{fig:test_acc_landscape}-\circled[0.6]{e}. They ranked latent dimensions to explain these yellow lights selected in \inlinep{figures/s2.pdf}, and found the dimensions of \textit{`DIM28'} and \textit{`DIM7'} explain the visual semantics of yellow lights with low confidence scores, as shown in \autoref{fig:test_acc_landscape}-\circled[0.6]{f}.

\vspace{-0.25em}
\subsubsection{How about potential failure cases?}\label{case_insight_rbt}
\vspace{-0.25em}
The domain experts were curious about how robust the detector was and where potential failure cases might come from. They examined the visual summary of robustness scores for all train objects, the lower part in \autoref{fig:teaser}-\circled[0.6]{b}. Overall, the detector had better robustness over green lights on left compared with the red lights on right in the view. Robustness was very low in the middle bottom part (selection \inlinep{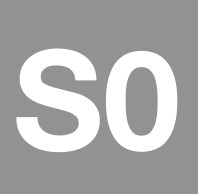} in \autoref{fig:teaser}-\circled[0.6]{b}). This part is mostly the dark and ambiguous traffic lights shown in the visual summary of training objects in \autoref{fig:teaser}-\circled[0.6]{b}. 

The domain experts then used latent dimensions to understand the low robustness area \inlinep{figures/s0.pdf}. As they selected the low robustness area and then ranked the dimensions, the top dimensions \textit{`DIM7'} and \textit{`DIM4'} capture the most important visual traits of the low robustness area (\autoref{fig:teaser}-\circled[0.6]{d}). It showed that the selected range of \textit{`DIM7'} represents color ambiguity and the one of \textit{`DIM4'} indicates dark color (\autoref{fig:teaser}-\circledtwo[0.6]{d}{1}).

They further examined the adversarial gradient directions of top dimensions in the low robustness area. They could see that the adversarial gradients push the visual appearance of traffic lights to move towards ambiguous appearance over the dimension \textit{`DIM7'} (the arrows over bars pointing into the center in \autoref{fig:teaser}-\circled[0.6]{d}). In the live detection view, \autoref{fig:teaser}-\circled[0.6]{c}, they also tried to put  various generated objects along the gradient direction to test detection results in the scene. 

%In the gradient direction view, \autoref{fig:teaser}-2c, they could see that the adversarial gradients push the visual appearance of traffic lights move towards the inside of the view, where the dark and ambiguous objects located.

They concluded that the detector was sensitive in the dark and ambiguous lighting conditions and therefore showed low robustness in these conditions. By examining the original scenes for low robustness cases, the lighting conditions were caused by some factors like far-away distance, shadow by building and tree or occluded by other objects.  
\vspace{-0.25em}
\subsubsection{Identifying noisy labels, confusing objects and more}
\begin{figure}[tb]
 \centering 

 \includegraphics[width=0.9\columnwidth]{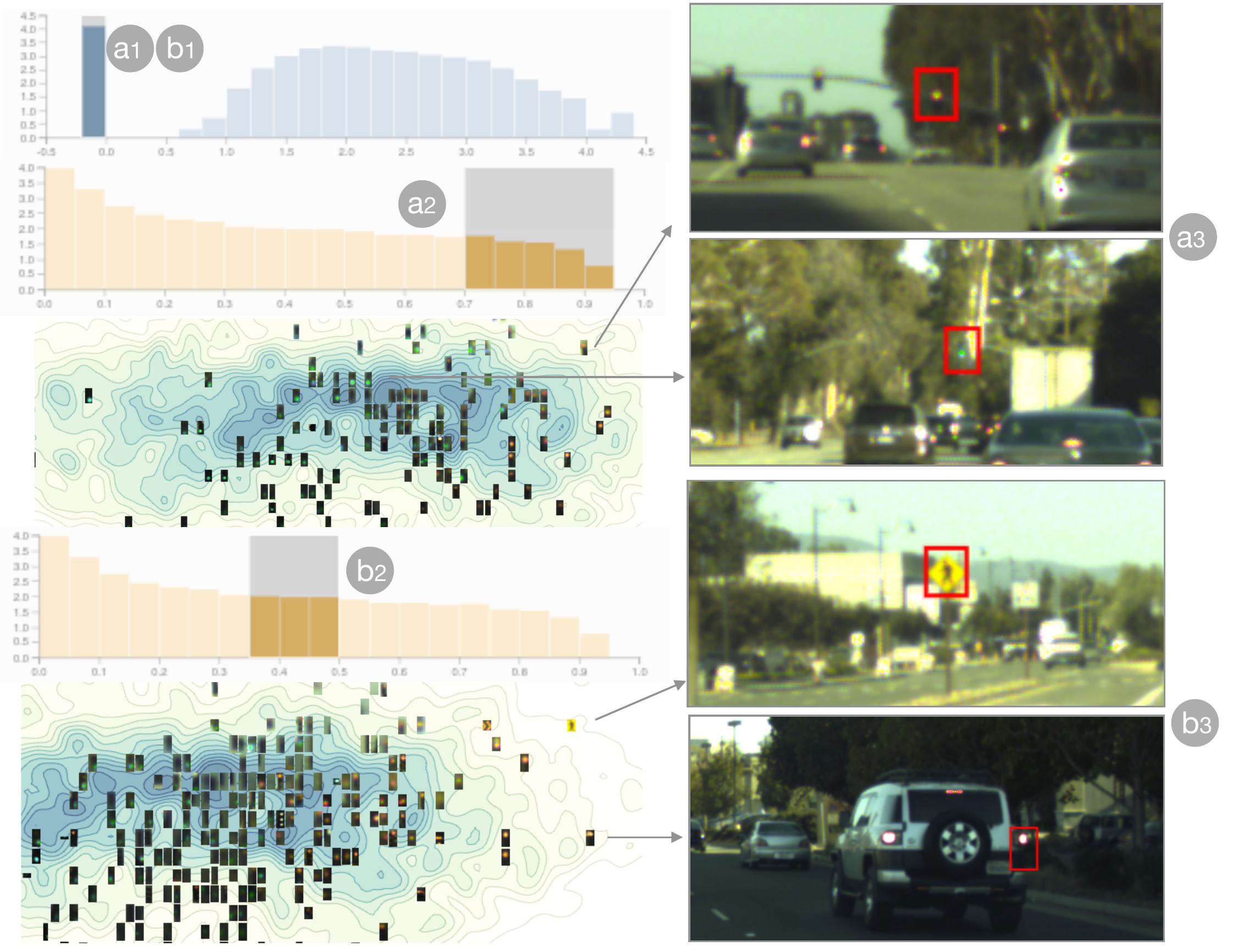}
 \vspace{-0.5em}
 \caption{ Diagnosis for data quality and confusing objects. (a) Identifying mislabeled data (a3) by selecting the predictions without ground truth (zero size predictions in a1) and having high confidence scores (a2);  (b) Understanding confusing objects (e.g. pedestrian sign and car rear lights in b3) by selecting the predictions without ground truth (b1) and having uncertain confidence scores (b2).}

 \label{fig:other_cases}
 \vspace{-1.5em}
\end{figure}

\vspace{-0.25em}
The multi-faceted visual analysis enabled the experts to conduct more useful diagnosis about data quality and detector behaviors. For example, it helped the experts identify data with missing labels, as shown in \autoref{fig:other_cases}-\circledtwo[0.6]{a}{3} by selecting \textit{FP}s with high confidence scores (\autoref{fig:other_cases}-\circledtwo[0.6]{a}{1}\circledtwo[0.6]{a}{2}). It also assisted users understand confusing objects for the detector, such as pedestrian sign and car rear lights in \autoref{fig:other_cases}-\circledtwo[0.6]{b}{3}, by selecting \textit{FP}s with uncertain confidence scores (\autoref{fig:other_cases}-\circledtwo[0.6]{b}{1}\circledtwo[0.6]{b}{2}). With various filtering combination in \autoref{fig:teaser}-\circled[0.6]{a}, the experts could conduct more analysis such as the impact of object size over performance.

\subsection{Visual Analytics Assisted Improvements}
\vspace{-0.25em}
With the insights from above analysis, domain experts developed three improvement strategies (\autoref{tab:aug_strategies}). These strategies aimed at generating more data from training data based on the \textit{VA} (Visual Analytics) insights, and augmenting model training to validate if we can improve the performance of either accuracy or robustness. 
\vspace{-0.25em}
\subsubsection{Distribution Guided Augmentation for Accuracy}
This strategy is to generate data from \textbf{training data} with the insight of shifted data distribution learned in \autoref{case_insight_acc} to augment the detector's accuracy (noted as \textbf{VA-Dist-Aug} in \autoref{tab:aug_strategies}). The experts already understood that there were under-representative traffic lights with bright red and yellow color in training data. 
%They also knew that the high brightness of red lights were controlled by dimension \textit{`DIM7'}, and \textit{`DIM13'}, and the yellow lights are controlled by dimension \textit{`DIM28'}, and \textit{`DIM7'}. 

In this strategy, for each case of under-representative bright red or yellow color, the selected dimensions (e.g. \textit{`DIM28'} and \textit{`DIM7'} for yellow lights) were used to generate more training data. The training objects in the selected ranges (e.g. $DIM28 \in [-3, -1.75]$ and $DIM7 \in [-0.75, 0]$) were first retrieved. For each retrieved object, $k$ ($k=5$ here) data points were uniformly sampled within the selected dimension ranges (e.g. $[-3, -2.75,...,-2]$ for \textit{`DIM28'} and $[-0.75, -0.6, ...,-0.15]$ for \textit{`DIM7'}). Then the sampled vectors were used to reconstruct object images. The generated object images are then blended back into original scene images as augmented data. 940 new objects were generated and used to fine tune the base detector along with original training data.  

\autoref{tab:exp_rslt_dist} shows accuracy improvement with five trials. The overall accuracy was improved from 0.478 to \textbf{0.493}, measured by $AP@IoU50$. The accuracy improvements over red and yellow lights were significant, although the green light accuracy had little drop. This indicates that the distribution guided data augmentation could help overall accuracy.

\begin{table}[tb]
 
  \caption{Data augmentation strategies assisted by visual analytics}
  \label{tab:aug_strategies}
  \includegraphics[width=\columnwidth]{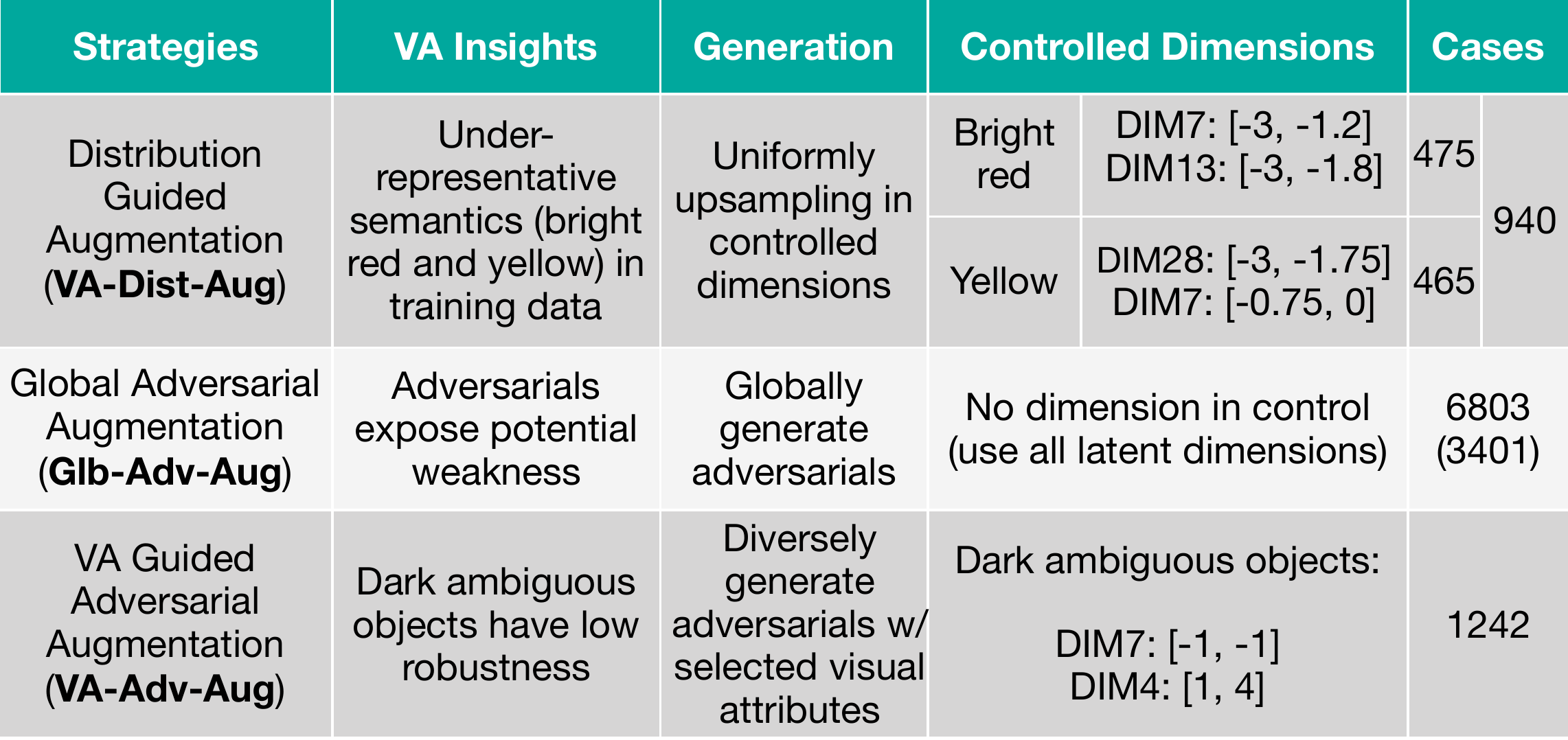}
\end{table}

\begin{table}[tb]
  \caption{Accuracy improvement with distribution guided data augmentation, measured by $AP@IoU50$ (5 trials, larger is better).}
  \label{tab:exp_rslt_dist}
  \includegraphics[width=\columnwidth]{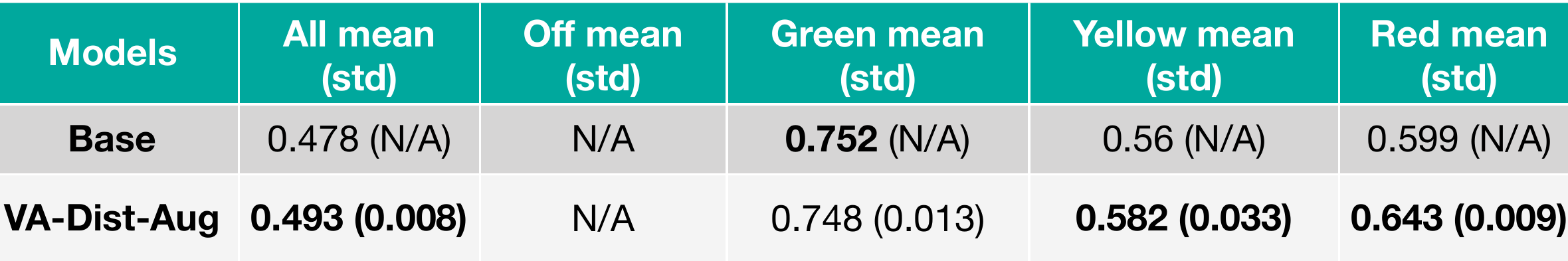}
  \vspace{-2em}
\end{table}

\begin{table}[tb]
 \vspace{-1em}
  \caption{Semantic robustness and accuracy improvements with adversarial augmentation, measuredby $AP@IoU50$ (5 trials, larger is better).}

  \label{tab:exp_rslt_adv}
  \includegraphics[width=\columnwidth]{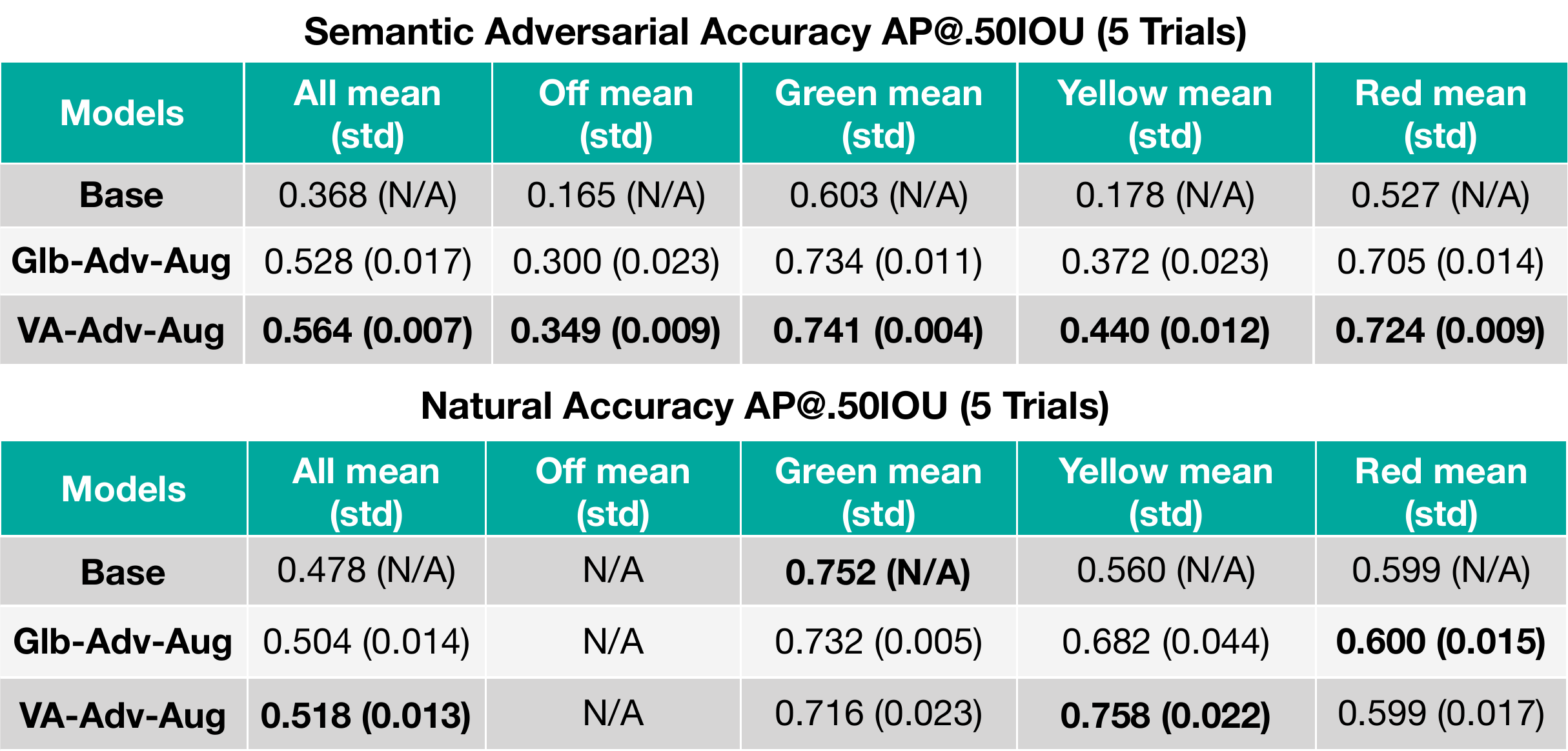}
  \vspace{-1em}
\end{table}

\vspace{-0.25em}
\subsubsection{Adversarial Guided Data Augmentation for Robustness}
\vspace{-0.25em}
A common way to improve robustness is adversarial training, namely training models with adversarial data \cite{Goodfellow2015}. With a similar idea, adversarial data were mixed with original training data and then used to fine tune the base model. To evaluate semantic robustness, the half of all adversarial objects was randomly selected as adversarial test data, and thus there was $\lceil 6803/2 \rceil =3402$ adversarial testing objects (along their original scene images). The second half adversarial objects ($3401$) were used for training and validation. 

Two adversarial guided data augmentation were experimented: \textbf{Glb-Adv-Aug} and \textbf{VA-Adv-Aug} in \autoref{tab:aug_strategies}. In the \textbf{Glb-Adv-Aug} strategy, all $3401$ adversarial objects were mixed with original training data for training and validation. The rationale for this strategy is that adversarials can expose potential weakness of a detector, and training with adversarials would help a model detect these adversarials. 

In the \textbf{VA-Adv-Aug} strategy, more adversarials were generated to augment model robustness with \textit{VA} insights that the dark ambiguous objects had low robustness. As the experts observed that the dark ambiguous objects were mostly explained by \textit{`DIM7'} and \textit{`DIM4'} (\autoref{fig:teaser}-\circled[0.6]{d}), they decided to generate 5 more adversarials for each object in the selected ranges of the two dimensions (i.e. $DIM7 \in [-1, 1]$, $DIM4 \in [1, 4]$). Also, they observed the adversarial gradients over \textit{`DIM7'} and \textit{`DIM4'} tends to push towards homogeneous dark and ambiguous directions, and thus they needed to diversify adversarial generation by fixing the gradients of these two dimensions. To further diversify the generation, other random 3 dimensions were also fixed during the process of adversarial searching, namely $ z_{ij}^{t+1} = z_{ij}^{0}, j \in [d4, d7, d_{rnd1}, d_{rnd2}, d_{rnd3}]$ in \autoref{algorithm:SemAdv}-\textit{line 10}. In this way, 1242 adversarial objects were obtained and mixed with global adversarial objects ($3401$) for adversarial training and validation. 

\autoref{tab:exp_rslt_adv} shows the performance improvement results with five trials. The results include two metrics of $AP@IoU50$ over adversarial test dataset for semantic robustness evaluation and the natural test dataset for accuracy evaluation. Both methods significantly improved semantic robustness with $AP@IoU50$ of \textbf{0.528} and \textbf{0.564} compared with base model of \textbf{0.368} over adversarial test data. Meanwhile, the two methods also improved the accuracy over the natural test data from \textbf{0.478} to \textbf{0.504} and \textbf{0.518}. This indicates that adversarial guided data augmentation can improve both accuracy and robustness performance.   

Moreover, the visual analytics guided method boosted both accuracy and robustness performance even further. With \textit{VA} insights , the \textbf{VA-Adv-Aug} enhanced the semantic robustness accuracy from {0.528} to \textbf{0.564} and the natural accuracy from {0.504} to \textbf{0.518} on top of the \textbf{Glb-Adv-Aug}. The robustness improvement across different categories (off, green, yellow and read) were also observed. This validates our assumption that the \textit{VA} insights can further improve the model performance of both accuracy and robustness with domain experts' knowledge.

% \begin{figure}[tb]
%  \centering 
%  \includegraphics[width=\columnwidth]{figures/adv_comparison.pdf}
%  \caption{Adversarial landscape comparison between base model (1) and adversarial augmented model (2). (3) shows the generated objects. }
%  \label{fig:adv_comparison}
% \end{figure}
\vspace{-0.25em}
\subsection{Domain Experts Feedback}
The system was deployed and used in the workplace of the domain experts. They also used the system to analyze in-production dataset and detectors. Here, we only reported the feedback that they had with above public data because of confidentiality issues. 

Overall, domain experts were impressed by the ``\textit{actionable insights}" obtained and also felt excited about experiment results on model improvement by ``\textit{injecting domain knowledge}" with little human-in-the-loop. For model validation, they found the tool’s capability to aid them in targeting weak spots is ``\textit{specifically useful}", and visual summary was ``\textit{the most useful feature}" by considering the data magnitude they have. They were able to be``\textit{swiftly aware about}" detectors’ sensitive regions with ``\textit{constrained adversarial attacks using the realistic augmentations}". For robustness interpretation, the tool enabled them understand network weaknesses with ``\textit{reasonable meaningful information}", such as gradient directions and semantic dimensions.

They also suggested improvements from practical perspectives, such as ``\textit{tracking quality trends}" over different datasets and models, and ``\textit{exporting and reporting}" actionable insights. Also, they expressed some visualizations were ``\textit{overwhelming}" to take improvement actions upon, such as gradient maps. This calls for us to strengthen our design principles of ``\textit{minimal human interaction}" and augmenting human cognition with ``\textit{interpretable representation learning}".

\section{Discussion and Future Work}
\subsection{What have we learned?}
\revisionblue{\textbf{Practical implications for detector evaluation in autonomous driving.} 
We learned some practical implications for both model evaluation and improvement in autonomous driving applications. One challenge in this domain is to deal with long-tail data distribution: although most cases could be covered by acquired dataset, rare cases comes from the long tail part \cite{Driverless2020}. To efficiently retrieve and collect data for model evaluation and improvement, we can search and retrieve data with potential issues via our approach (e.g. high brightness and dark ambiguous objects for detectors), from a huge candidate dataset, or we can collect new driving scenes with such lighting conditions to improve models. This data-driven visual analytics approach bears much potentials for model understanding and improvement.} 

%For example, https://www.youtube.com/watch?v=Q0nGo2-y0xY
%https://www.forbes.com/sites/chunkamui/2018/02/28/driverless-cars-90-percent-done-90-percent-left-to-go/

\textbf{Augmenting human cognition with representation learning.}
Semantic representation learning shows promising capability of augmenting human cognition to understand complicated data and model space of deep neural networks. Some learning approaches, such as feature maps\cite{Szegedy2014}, concepts and feature attributions\cite{Bau2017}, directly extract interpretable representations from model spaces. In this work, semantic representation learning extracts human-friendly representations from data space, and then reveal feature contribution towards model predictions. Both approaches serve as more powerful technologies of understanding, interacting with and improving AI systems \cite{carter2017using}. 

\revisionblue{\textbf{Minimal human-in-the-loop and maximizing actionable insights.} 
We aim at utilizing \textit{minimal human interactions} \cite{Carroll1990} to inject domain knowledge into models during system design. Towards this end, two learning components of semantic representation and adversarial generation are introduced to complete heavy-lifting work: semantic representation extracting interpretable dimensions for human-friendly visualization and interaction, and adversarial generation reducing searching space to probe model potential weakness. Meanwhile, we also target maximizing actionable insights with minimal interaction (e.g. Rank-To-Interpret feature), but also provide rich interactive visualization to enable in-depth exploration on demand. However, it is a challenging task to balance the level of details (e.g. more details for adversarial searching) to be presented so that users can make sensible and reliable decisions with few interactions \cite{shneiderman}. This calls for more investigation. }

\subsection{Limitations and future work}
\textbf{Extended to multi-object detection.} Our current system aims at traffic light detection problem, namely a single object detector. However, the framework is also applicable to the problem of multi-object detections (e.g. person, car, truck, bike, rider, sign, etc). We plan to extend this wok to generic object detectors for autonomous driving.  

\textbf{Representation learning with powerful expressiveness.} 
%Semantic representation learning serves as a cornerstone to interpret, assess and improve model performance for this work. 
In spite of promising semantic interpretation in disentangled representation learning \cite{Burgess2018}, we need non-trivial visual design, such as \hpcp\, to associate semantics with latent dimensions (e.g. which dimensions explain object color or brightness). One possible remedy is to investigate representation learning methods of combining neural networks and symbolic learning \cite{Besold2017, Wang2020}, by associating symbolic schema (e.g. decision tree, or rules) with latent representations.

\textbf{Understanding detector beyond objects.}  The current method largely uses the visual semantics of objects to understand and improve object detectors, but leaves the context of driving scenes for future work. 
%The rationale of only using object information is built upon the observation that the mainstream detectors only focus on the proposed regions without considering the scene context. The context may provide additional information to help detection (e.g., traffic lights may have higher chance appearing at street vanishing points). 
We can extend this approach by parsing driving scene semantics with additional representation learning components \cite{zhao2017pyramid}.

\section{Conclusion}
In this work, we propose a visual analytic system, \systemname, to assess, understand, and improve the accuracy and robustness of traffic light detectors for autonomous driving. This approach is built upon a representation learning to augment human cognition with human-friendly visual summarization, and a semantic adversarial learning to expose interpretable robustness issues with minimal human interactions. We also demonstrate the effectiveness of performance improvement strategies derived with \systemname, and illustrate practical implications for real-world problems in production environments. We hope this work can capture a silver of ways of applying the approaches of visual analytics and human-in-the-loop to alleviate some trustworthy AI issues in autonomous driving domain.

% The authors wish to thank A, B, and C. This work was supported in part by
% a grant from XYZ (\# 12345-67890).}

%\bibliographystyle{abbrv}
\bibliographystyle{abbrv-doi}

\bibliography{main}
\end{document}